\pdfoutput=1
\documentclass[article]{elsarticle}
\usepackage{hyperref}

\usepackage{graphicx}
\usepackage{array}
\usepackage{amssymb}
\usepackage{arydshln}
\usepackage{setspace}
\usepackage[pdf]{pstricks}
\usepackage{multirow}
\usepackage{subcaption}
\newcolumntype{P}[1]{>{\raggedright\arraybackslash}p{#1}}
\usepackage[ruled,norelsize]{algorithm2e}

\begin{document}

\begin{frontmatter}

\title{PRNU Based Source Camera Attribution for Image Sets Anonymized with Patch-Match Algorithm}

\author{Ahmet Karak\"{u}\c{c}\"{u}k\fnref{akfootnote}}
\author{A. Emir Dirik\fnref{aedfootnote}*}
\address{Uluda\u{g} University, Nil\"{u}fer, Bursa, Turkey}
\fntext[akfootnote]{Dept. of Electrical-Electronics Engineering.}
\fntext[aedfootnote]{* Corresponding Author. Dept. of Computer Engineering. edirik@uludag.edu.tr}

\begin{abstract}
Patch-Match is an efficient algorithm used for structural image editing and available as a tool on popular commercial photo-editing software. The tool allows users to insert or remove objects from photos using information from similar scene content. Recently, a modified version of this algorithm was proposed as a counter-measure against Photo-Response Non-Uniformity (PRNU) based Source Camera Identification (SCI). The algorithm can provide anonymity at a great rate (97\%) and impede PRNU based SCI without the need of any other information, hence leaving no-known recourse for the PRNU-based SCI. In this paper, we propose a method to identify sources of the Patch-Match-applied images by using randomized subsets of images and the traditional PRNU based SCI methods. We evaluate the proposed method on two forensics scenarios in which an adversary makes use of the Patch-Match algorithm and distorts the PRNU noise pattern in the incriminating images he took with his camera. Our results show that it is possible to link sets of Patch-Match-applied images back to their source camera even in the presence of images that come from unknown cameras. To our best knowledge, the proposed method represents the very first counter-measure against the usage of of Patch-Match in the digital forensics literature.
\end{abstract}


\end{frontmatter}


\section{Introduction}

Photo Response Non-Uniformity Noise (PRNU) was found to be very valuable in source camera identification (SCI) since its introduction into the forensics literature \cite{Lukas2005a,Lukas2005b,Lukas2005c,Lukas2006b,Goljan2007,Goljan2009a}. It is used by many agencies for identifying the origin devices of digital images. Robustness of this method was also evaluated against many edge cases \cite{Goljan2012,Gloe2012b} and was even experimented on topics beyond the scope of image forensics \cite{Valsesia2017}. Researchers have also looked to improve the handling and querying PRNU fingerprints, from compression \cite{Goljan2011a,Bayram2012} to fast search algorithms \cite{Taspinar2018}. However, as in many forensics \& security research, counter measures against PRNU based SCI were also developed from cloning \cite{Goljan2011}, to denoising \cite{Dirik2014d,Karakucuk2015} attacks. In many of these counter measures, researchers assumed the knowledge of the underlying methodology of the PRNU fingerprint estimation and detection prior or during the attack. In contrast, image modification techniques based on image content, such as seam-carving (which alters image aspect-ratio) and re-alignment (i.e. panorama) were also considered for the purposes of anonymization \textcolor{black}{as they distort the spatial synchronization of the PRNU pattern and} were shown to increase the computational cost of SCI \cite{TIFS_dirik2014,Karakucuk2015SPIE,Taspinar2017}. 

From the perspective of an adversary, the real advantage of such methods are their blind applicability. However, alteration of the content and form of images might not be desirable. 
\textcolor{black}{A structural image editing algorithm called ``Patch-Match'', in contrast, does not alter the form of the image and the image content is mostly preserved, which makes it a suitable tool to de-synchronize the PRNU pattern, and to use against the PRNU based SCI} 
was first reported in 2016 \cite{Entrieri}. Using this method, the authors efficiently redistributed the pixels of image patches to produce shuffled, but good looking images. This can be done without neither any prior evaluation of potential detection schemes nor any information related to the camera. As a result, the source cameras of patch-matched images are hardly identifiable. This notion gives any adversary the ability to become anonymous against PRNU based SCI using the Patch-Match algorithm as a low-cost, ``one-click solution''. 

In this paper, we evaluate a strategy that can be adopted against Patch-Match based PRNU counter-forensics attack. \textcolor{black}{Since the PRNU pattern becomes very distorted after this attack}, the individual Patch-match-applied images could not be linked to their PRNU fingerprints directly. \textcolor{black}{However,} our studies have shown that, such images can be grouped randomly into small subsets, and subsets having the majority of images from a questioned camera can be determined successfully with the proposed strategy. Specifically, we would like to answer the following questions regarding this attack:
\begin{itemize}
\item Can we verify the source camera device of a set of Patch-Match-applied images taken with the same camera device?
\item Can we identify the source camera device of images anonymized with Patch-Match algorithm in a mixed image set comprising images taken with two different cameras?  
\end{itemize}
To answer these questions, we have simulated two different scenarios and conducted experiments to evaluate the performance of the proposed approach.

Our analysis in this paper have been conducted on the dataset cited in \cite{Korus2016TIFS, Korus2016WIFS}. The patch-match attack implementation is the one mentioned earlier in this paper \cite{Entrieri}. Interested readers can also access the Patch-match-applied version of the dataset used in this study here: \href{https://github.com/akarakucuk/2019_PM_SCI_DATA/}{github.com/akarakucuk/2019\_PM\_SCI\_DATA/}.

\section{Photo-Response Non-Uniformity Based Source Camera Identification and Patch-Match Algorithm}

In this section, we are going to introduce the PRNU based Source Camera Identification scheme and the Patch-Match algorithm very briefly. 
In the next section, we outline the the proposed method to identify subsets of images processed by the Patch-Match algorithm.

\subsection{PRNU Based Source Camera Identification: The conventional method}
In a camera sensor, photo-sites' response $l$ to a photon intensity $l_\mathrm{0}$ varies as a result of the imperfections in manufacturing process which is called as Photo-Response Non-Uniformity Noise (PRNU). These variations generates a noise pattern (PRNU fingerprint) denoted by $\mathrm{\mathbf{F}}$ that was proportional to the size of a sensor and serves as a attributable link to a particular imaging sensor. Extraction of PRNU fingerprint could be explained through imaging sensor output model used in \cite{Chen2007,Goljan2009a} with matrix notation: $\mathrm{\mathbf{L}}=\mathrm{\mathbf{L}}_\mathrm{0}+\mathrm{\mathbf{L}}_\mathrm{0} \mathrm{\mathbf{F}}+\mathrm{\mathbf{\Gamma}}$ where $\mathrm{\mathbf{\Gamma}}$ represents other, mostly-random noise sources and $\mathrm{\mathbf{L}_0}$ represents all intensity values apparent to the sensor for a still image. The PRNU fingerprint $\mathrm{\mathbf{F}}$ could be estimated from a number of wavelet noise residues $\mathrm{\mathbf{W}}_{1},...,\mathrm{\mathbf{W}}_{n}$ \cite{Lukas2006b}, s.t. $\mathrm{\mathbf{W}} = \mathrm{\mathbf{L}}-\mathrm{denoiser}(\mathrm{\mathbf{L}})$, using the MLE estimator shown in \cite{Chen2007d}:

\begin{equation}
\label{eq:fingergeneration} \hat{\mathrm{\mathbf{F}}}=
\frac{\sum_{i=1}^{n}\mathrm{\mathbf{W}}_{i} \mathrm{\mathbf{L}}_{i}} {
\sum_{i=1}^{n}\mathrm{\mathbf{L}}_{i}^2}
\end{equation}

and in \cite{Goljan2009a} and then be used to find similarity between a noise extract $\mathrm{\mathbf{W}}_i$ of a query image $\mathrm{\mathbf{L}}_i$ and MLE-estimated PRNU pattern, $\hat \mathrm{\mathbf{F}}$ with peak-to-correlation energy (PCE) $\rho=\mathrm{PCE}(\mathrm{\mathbf{W}}_i,\mathrm{\mathbf{L}}_i \hat \mathrm{\mathbf{F}})$ which uses normalized correlation operator between the residue and the PRNU fingerprint with notable modifications \cite{Goljan2009a}.

\subsection{PRNU De-synchronization Attack by Patch-Match}

Patch-Match is an algorithm used for in-painting of images. It works by computing a dense neighborhood field of image patches with a pre-defined size, and uses information to match and replace such patches. Commercially available implementations of the method can exchange such blocks between images of multiple scenes at almost real-time. The algorithm can be forced to a single image to insert or remove contents. The PRNU de-synchronization attack implementation of it also imposes restrictions to avoid matching of a block by itself \cite{Entrieri}, and applies additional filtering to avoid significant degradation of image quality.

As Patch-Match shuffles a given image with its most similar patches, it implicitly breaks the synchronization between the noise residue of a Patch-Match-applied image and a PRNU fingerprint of the camera that took the image, by distorting the spatial correspondence between the noise residue and the PRNU pattern. 

This gives the Patch-Match based attack the advantage of blind applicability, as it requires neither an analysis nor any prior information other than the image it is being applied to. This advantage makes the method very versatile to the conventional PRNU based SCI approach. In Figure \ref{fig:examplePMimages}, a few examples of Patch-Match-applied images can be seen along with image quality levels in terms of PSNR.

\newcommand{\mysize}{0.505}
\newcommand{\arabosluk}{\vspace{.1px} } 

\begin{center}
\begin{figure}
\captionsetup{size=small}
\captionsetup[subfigure]{justification=centering}

\begin{subfigure}{0.23\textwidth}
\includegraphics[trim = 849px 469px 899px 439px, clip, scale=\mysize]{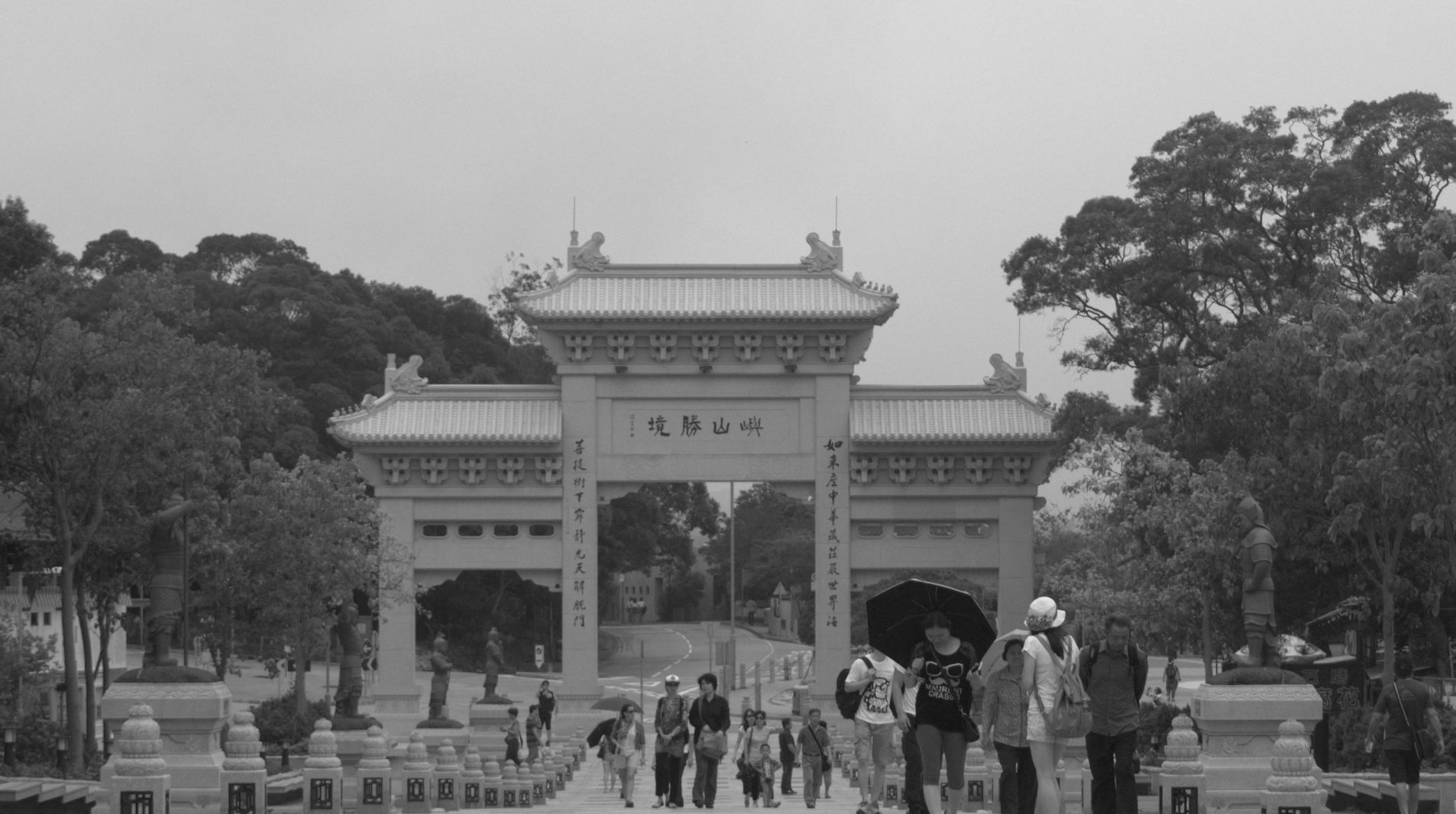}
\caption{PCE=2646} \label{subfig:woPM1}
\end{subfigure} 
\arabosluk
\begin{subfigure}{0.23\textwidth}
\includegraphics[trim = 706px 858px 1042px 50px, clip, scale=\mysize]{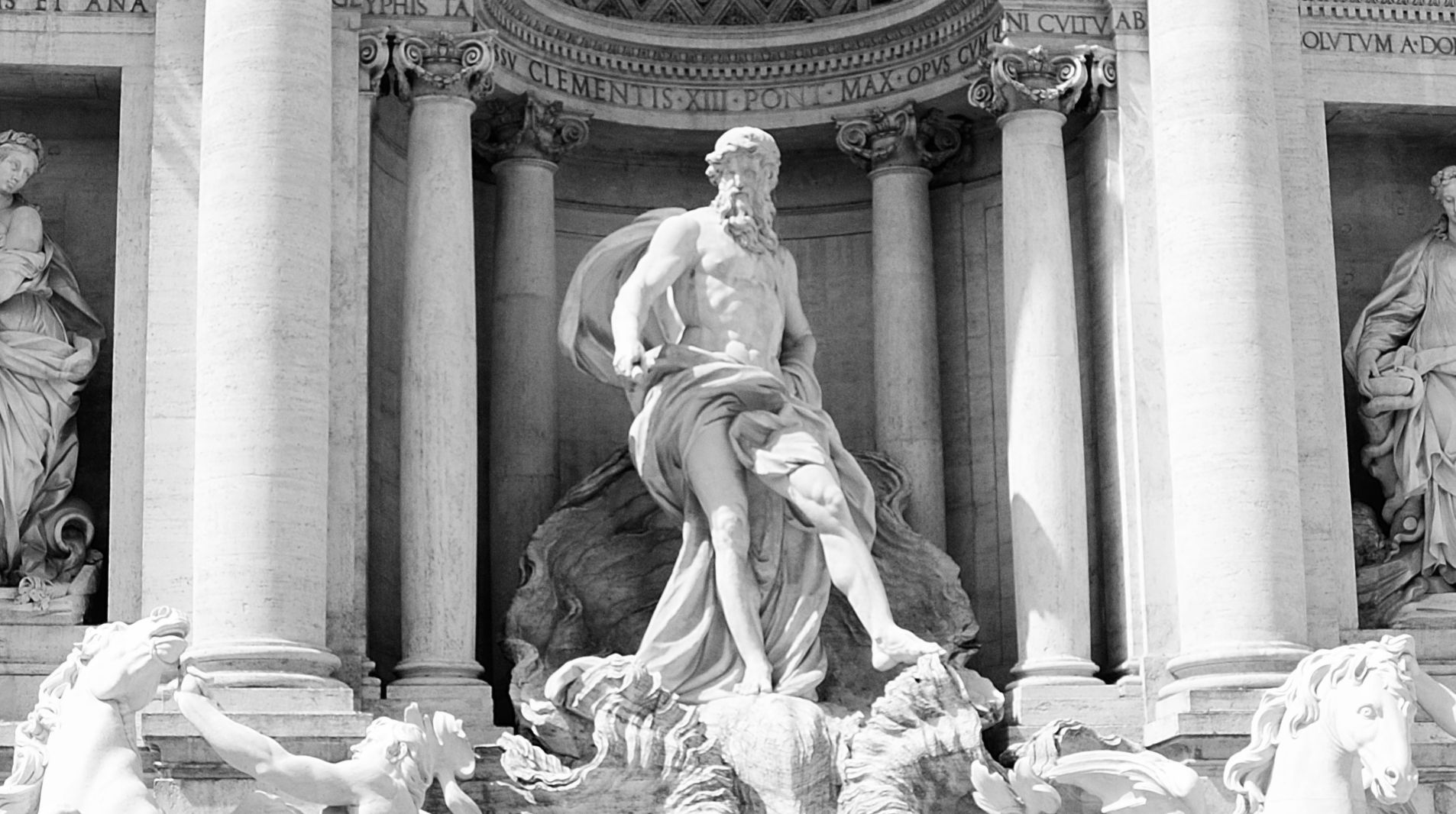}
\caption{PCE=1531} \label{subfig:woPM2}
\end{subfigure}
\arabosluk
\begin{subfigure}{0.23\textwidth}
\includegraphics[trim = 906px 519px 842px 389px, clip, scale=\mysize]{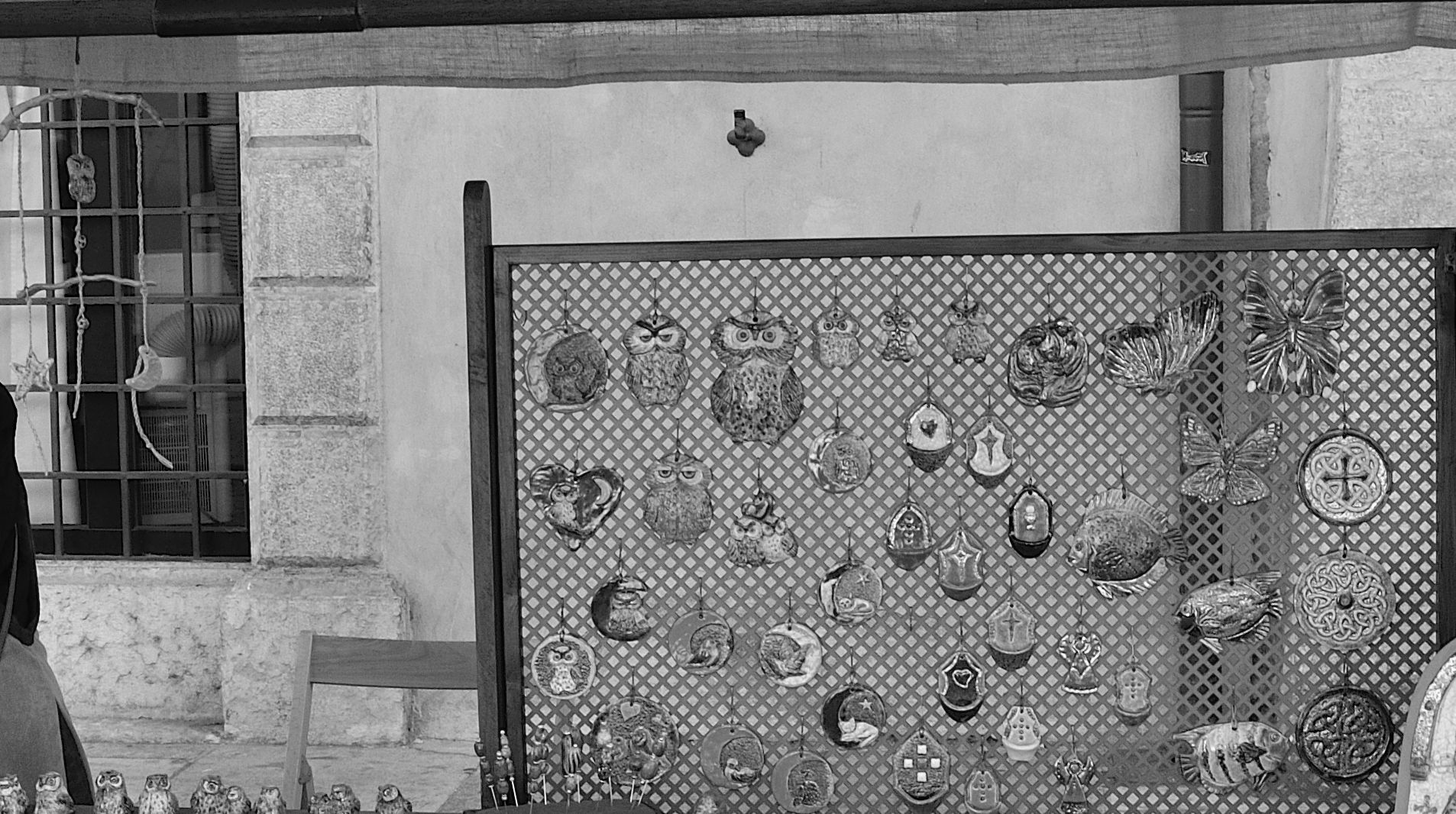}
\caption{PCE=872} \label{subfig:woPM3}
\end{subfigure} 
\arabosluk
\begin{subfigure}{0.23\textwidth}
\includegraphics[trim = 756px 569px 992px 339px, clip, scale=\mysize]{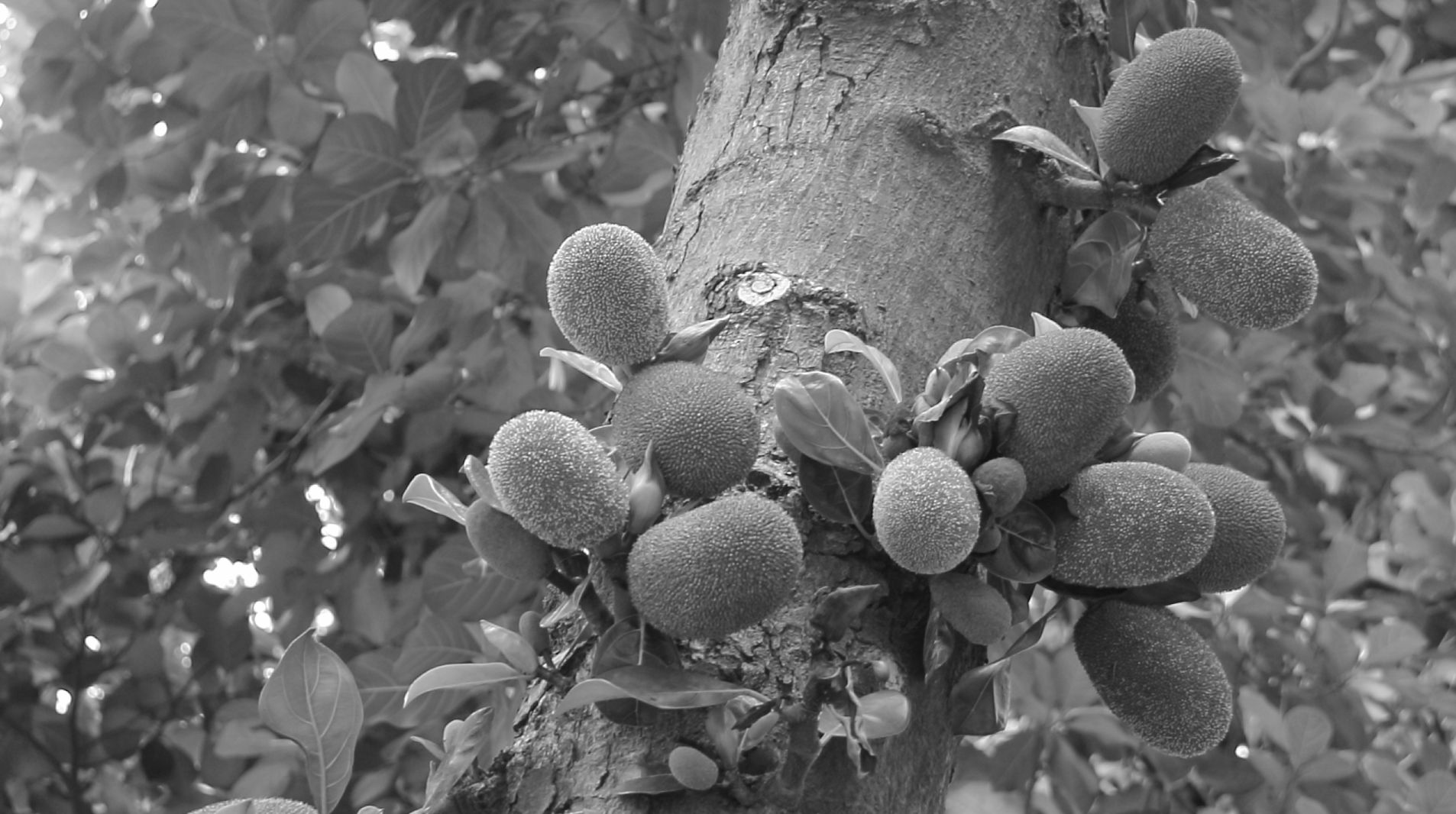}
\caption{PCE=1502} \label{subfig:woPM4}
\end{subfigure}

\bigskip
\begin{subfigure}{0.23\textwidth}
\includegraphics[trim = 849px 469px 899px 439px, clip, scale=\mysize]{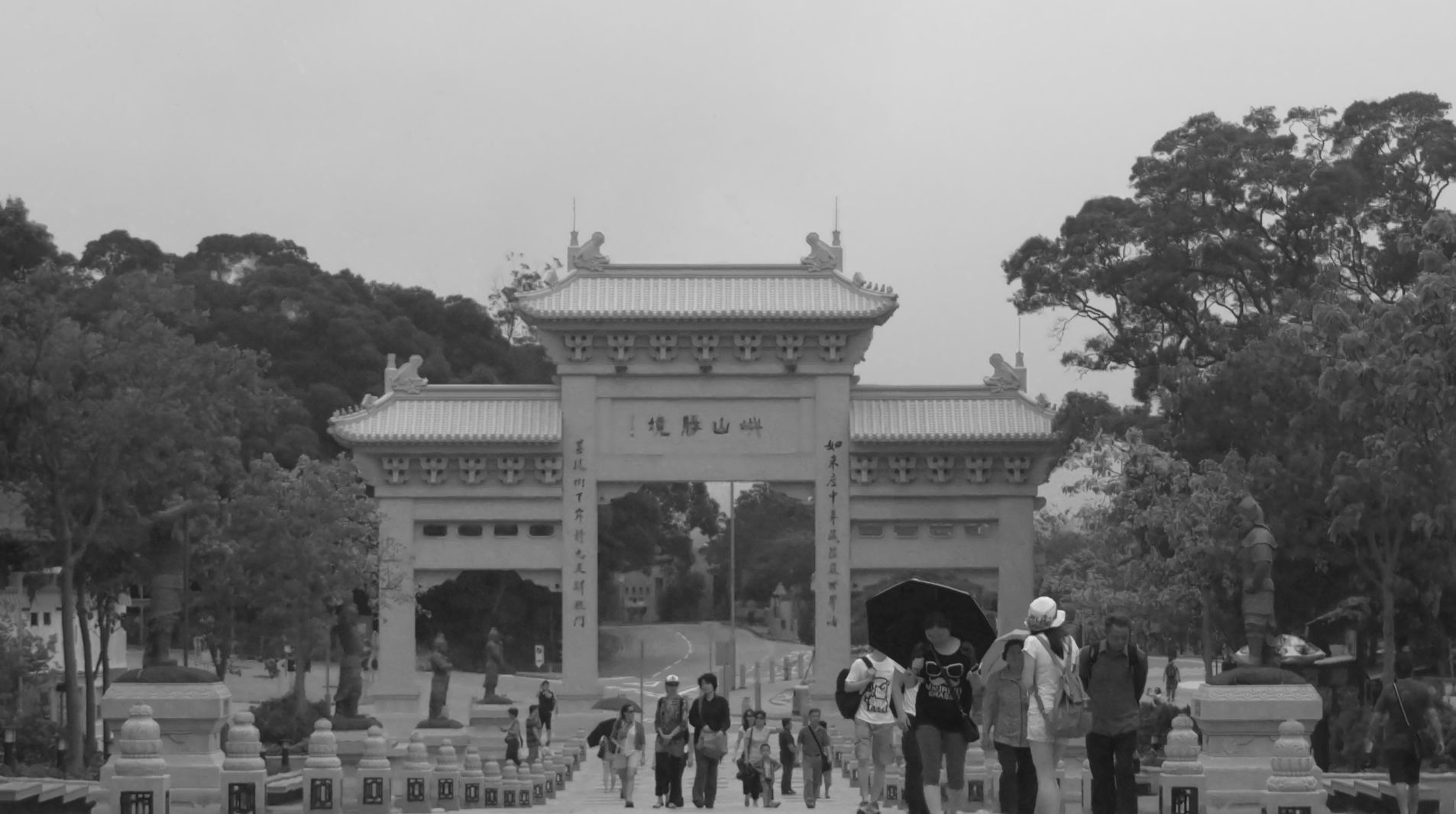}
\caption{PCE=5.6\\36dB,MPR=83\%} \label{subfig:wPM1}
\end{subfigure} 
\arabosluk
\begin{subfigure}{0.23\textwidth}
\includegraphics[trim = 706px 858px 1042px 50px, clip, scale=\mysize]{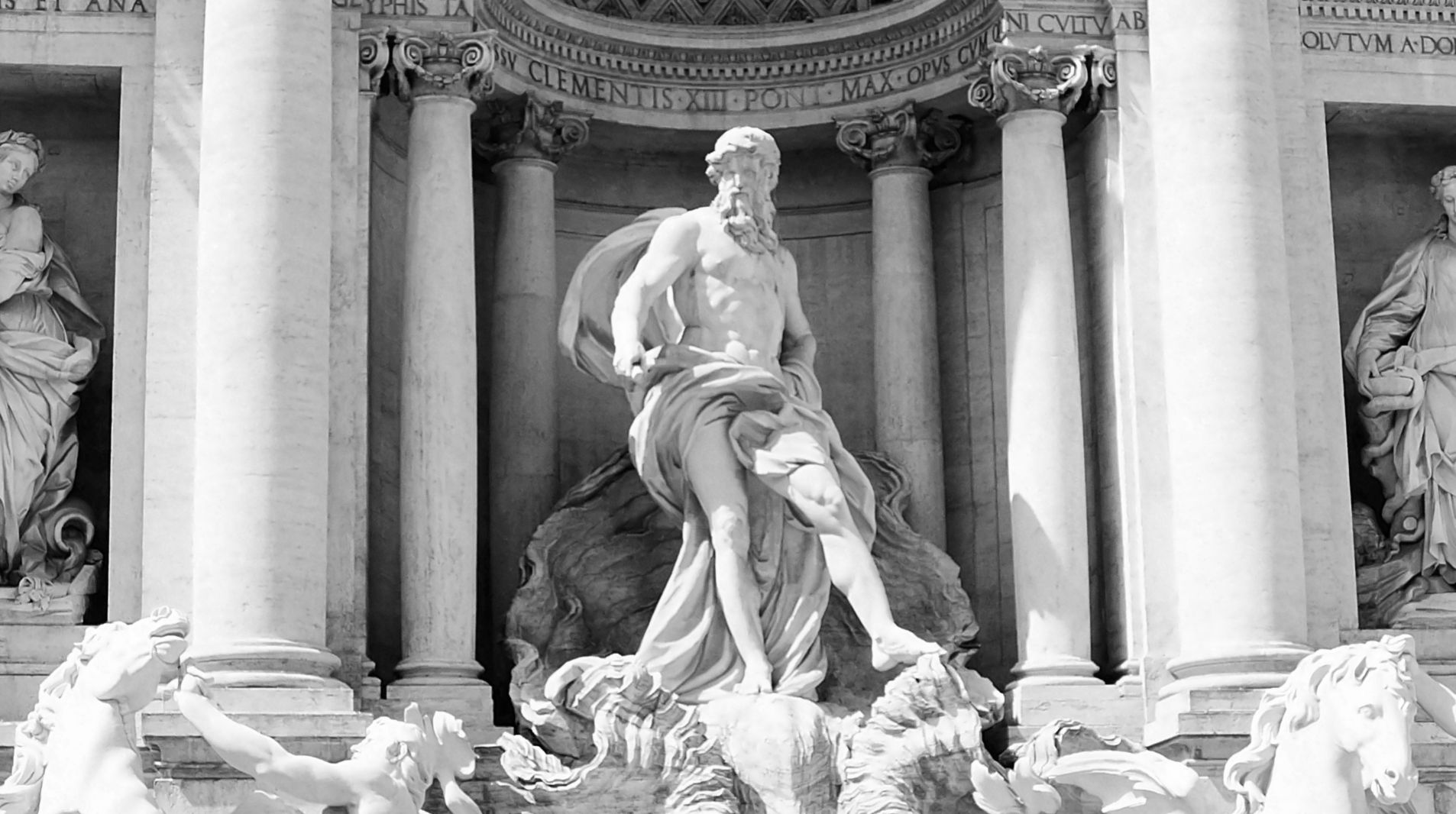}
\caption{PCE=0.2\\35dB,MPR=86\%} \label{subfig:wPM2}
\end{subfigure}
\arabosluk
\begin{subfigure}{0.23\textwidth}
\includegraphics[trim = 906px 519px 842px 389px, clip, scale=\mysize]{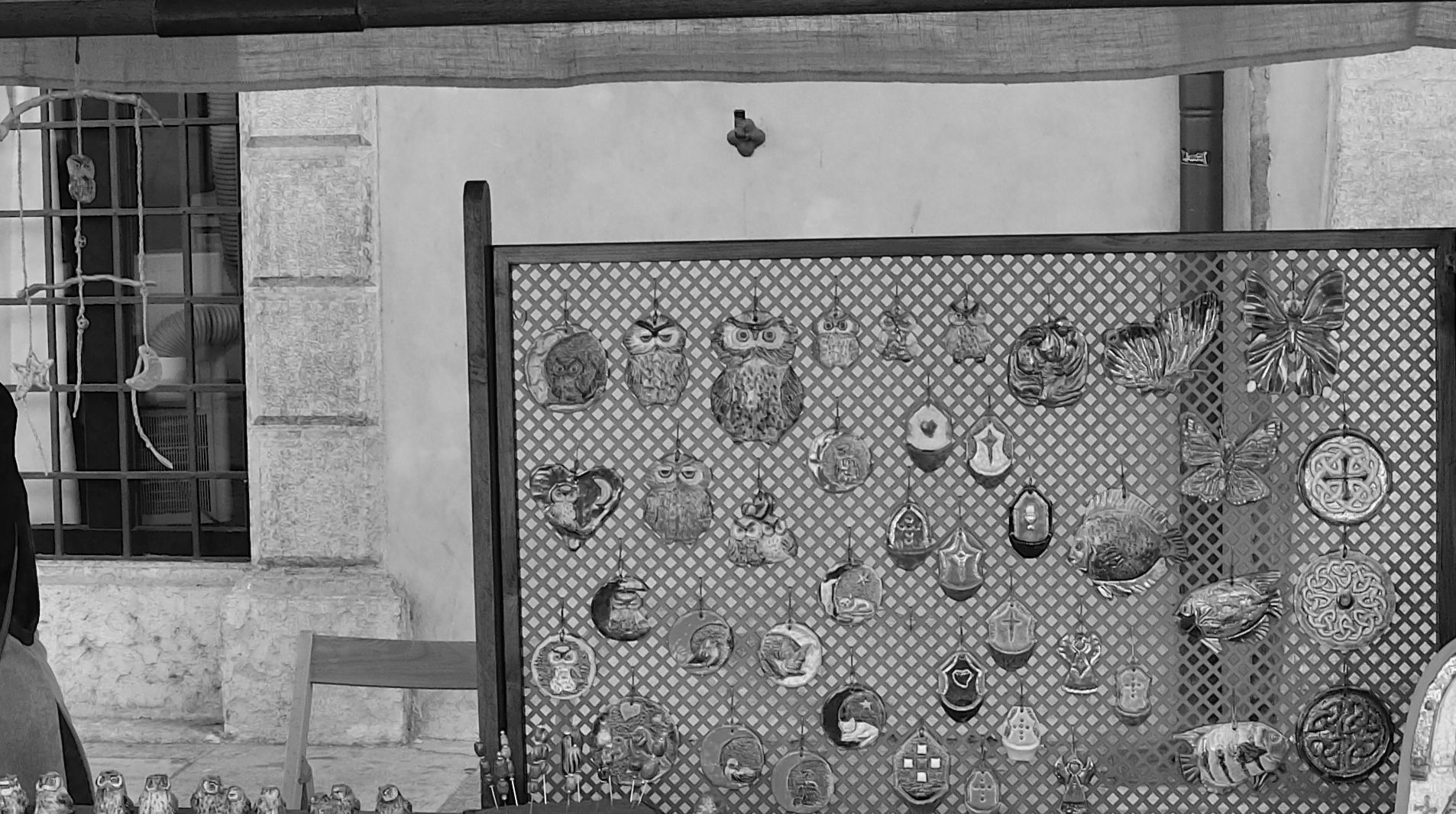}
\caption{PCE=4.8\\30dB,MPR=90\%} \label{subfig:wPM3}
\end{subfigure} 
\arabosluk
\begin{subfigure}{0.23\textwidth}
\includegraphics[trim = 756px 569px 992px 339px, clip, scale=\mysize]{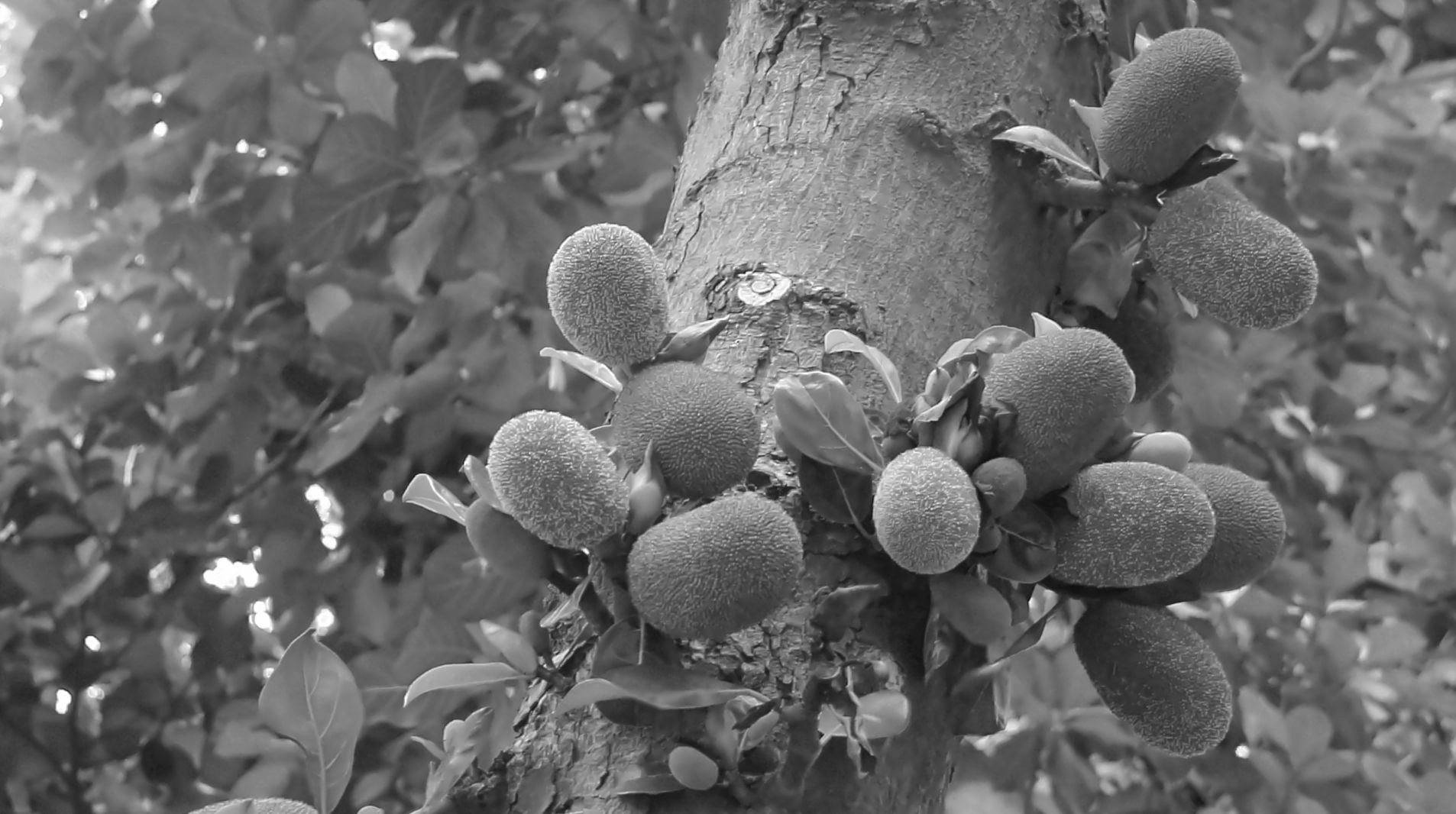}
\caption{PCE=-0.9\\35dB,MPR=84\%} \label{subfig:wPM4}
\end{subfigure}

     \caption{Example images. Images in the first row are original, whereas those in the second row are PM-images. In addition, the first column's images are from camera A57, the second's from D7000, the third's from D90 and the last column's are images from 60D. Under each PM-image, PSNR in terms of dB and MPR values are also noted. For MPR description and equation, please see Section \ref{section:theproposedmethod} and Eq. \ref{eq:mpr}. Example images are trimmed to $158\times158$ pixels in size, with half scale for better viewing.}
     \label{fig:examplePMimages}
\end{figure}
\end{center}

The Patch-Match-applied images tend to become flatter, and lose some fine details (such as the thin lines and small spots) w.r.t. their original counterparts. The example images in Figure \ref{fig:examplePMimages} show the effect of Patch-Match algorithm. Reduction in PCE value is evident, and images still have acceptable image quality. Please note that both versions are re-compressed only once and trimmed from the very same coordinates.

On the use of terminology, we would like to indicate that in the rest of the paper we use ``PM'' as a short-hand to refer to the Patch-Match algorithm, and ``PM-images'' to indicate images that have been processed with the PM algorithm. More information on the use of remaining notation will be provided in the coming sections.

\section{Source Attribution of Images Anonymized with Patch-Match Algorithm}
\label{section:theproposedmethod}
To be able to gain SCI opportunity on PM-images, we looked for weaknesses of the PM from the perspective of a forensic analyst. The PM attack exploits the redundancies (similarities) within a given image and shuffles small portions of images with other, same sized portions. The redundancies here are specific to each scene content, thus they can be assumed to be randomly distributed between different scenes. 

This gives us an initial idea for identifying the sources of PM-images, by incorporating multiple PM-images, we can attribute these image combinations to their sources. 
Our next observation is regarding the differences between a PM-image and its source. 

That is, on each image we evaluated, the \textcolor{black}{ratio} of pixels that stood the same was around 15\% on average, and fluctuated between 55\% to 8\%. This indicates that it may be possible to attribute the source of PM-images to their originating camera, however it may require more than one image to do so, in other terms, we can merge the noise residues from PM-images, then check for similarities between this ``merged" noise residue 
and the PRNU fingerprint of the analyst which is assumed to be taken from pristine images of the query camera. For all experiments, we assume that the analyst is able to estimate a PRNU fingerprint from 25 pristine images from the query camera and denote this estimate with a bold symbol $\mathrm{\mathbf{F}_q}$.

The adversary, on the other hand, has applied several countermeasures against SCI: PM and meta-data removal on any incriminating image he has. Having received a storage media full of incriminating images which have unknown origin, and a query camera, the analyst's task is seemingly simple: Finding out if any incriminating image on this disk was obtained with the query camera.

The analyst estimates the PRNU fingerprint of the query camera, $\mathrm{\mathbf{F}_q}$. Using this estimate, she tries to attribute the incriminating images on a storage media with the query camera using the classical PRNU based SCI methods. As all images originating from the query camera were PM-images, this attempt fails. However, using the proposed method, she can try to see if combining subsets of these images would increase his/her chances to attribute these subsets of images to the query camera. As our results will show, using the proposed method, she will have more chances to link the image subsets with at least one PM-image from the query camera, and in some cases has 100\% chance to find subsets of images which have at least one PM image from the query camera. 

Briefly, we'll consider two scenarios: Scenario \#1 shows when the storage media contains only PM-images that originate from the query camera. The disk content for the Scenario \#2 on the other hand, is mixed and has three types of images, with the following ratios: i) 50\% of the total disk content is from PM-images of the query camera; (ii) 25\% PM images that originate from the unknown camera, and lastly (iii) 25\% original images from the unknown camera. Images used in the scenario \#1 will be called $\mathrm{S_\alpha}$, and those used in Scenario \#2 will be called $S_\mathrm{\Sigma}$.

\begin{figure}[!t]
  \begin{algorithm}[H]
   \caption{Pseudo-code for PRNU Subset and Fusion SCI using Small Image Sets}
   \emph{inputs:}\\
   $\textrm{X} \gets$ list of all images in the storage media\;
   $K \gets$ initialize the number of \textcolor{black}{subsets} to 100\;
   $\tau \gets$ initialize PRNU similarity threshold to 50 in terms of PCE\;
   $n \gets$ \textcolor{black}{subset} length\;
   $q \gets$ query camera label\; 
   $\mathrm{\mathbf{F}_q} \gets$ load the PRNU fingerprint of query camera\;
   $o \gets$ unknown camera's label\; 
   $p \gets$ corresponding Case ID from $q$ and $o$\;
   $\mathrm{S_\alpha} \gets$ list of PM-images from query camera ($q$)\;
   $\mathrm{S_\beta} \gets$ list of PM and non-PM-images from unknown camera ($o$)\;
   $\Phi\gets {\varnothing}$ initialize an empty fusion set\;
    $\mathrm{S} =\left \{ \textrm{X}_{1}, ... , \textrm{X}_{N} \right \} \in \{ \mathrm{S_\alpha} \cup \mathrm{S_\beta} \} \gets $ populate image set (file paths)\;

    \emph{iteration:}\\
       
   \For{$k := 1$ to $K$}
   {
      $\mathrm{S}_{k} \gets$ A new subset of $n$ randomly selected images from $\mathrm{S}$ \textcolor{black}{within the loop according to Eq. \ref{eq:subsetcondition}}\;
      $\mathrm{\mathbf{F}}_k\gets$GenerateFingerprint($\mathrm{S}_{k}$)\;
      ${\rho}_{k} \gets$ PCE value between $\mathrm{\mathbf{F}}_{k}$ and $\mathrm{\mathbf{F}_q}$\;
     \If {${\rho}_{k} \geq \tau$} 
     {$\Phi_p \gets \Phi_p  \cup \{\mathrm{S}_{k}\} $, add the set of image paths into fusion set\;
     \textbf{return} $k$th subset ($\mathrm{S}_{k}$) and PCE value ($\rho_k$)\;}      
   }
   \If {$\Phi_p$ is not empty} {
   $\mathbf{\mathcal{F}}_{p} \gets$ GenerateFingerprint($\Phi_p$)\;
   $\varrho_{p} \gets$ PCE value of $\Phi_p$\;
   \textbf{return} $p$th fusion set ($\Phi_p$) and PCE value (${\varrho}_{p})$\;
   }
  \end{algorithm}
\end{figure}

In this section, for the purpose of simplicity, we are going to denote a ``pseudo-operator'' denoted with $\mathrm{GenerateFingerprint(\mathrm{S})}$ to describe a PRNU fingerprint generation function. This function accepts any list of images $\mathrm{S}$, s.t. $\mathrm{S=\{image1.png,image2.png,...\}}$, and it is essentially a wrapper for the process of PRNU fingerprint estimation, as shown in Eq. \ref{eq:fingergeneration} thus produces a PRNU fingerprint estimate from any given set of image list $\mathrm{S}$, such that  $\mathrm{\mathbf{F}_S}=\mathrm{GenerateFingerprint(S)}$. Whenever we want to specify a specific type of images, we add a subscript next to the set, e.g. $\mathrm{S_\alpha}$. 

The subscripts denote the origin of the image set, where $\mathrm{\alpha}$ represents the PM-images from the query camera (with camera PRNU fingerprint known to the analyst), $\mathrm{\beta}$ represents both PM and pristine images from unknown cameras (whose PRNU fingerprints are unknown to the analyst). If, however, both two collections are used, the subscript $\Sigma$ is used. Similarly, whenever we say, ``PCE of $\mathrm{S}$'' we refer to a PRNU similarity value in terms of PCE, between a PRNU fingerprint generated from a set $\mathrm{S}$, and the query camera PRNU fingerprint $\mathrm{\mathbf{F}_q}$ the analyst has, 
\begin{equation} \label{eq:pseudoR}
\textrm{``PCE of S''} = \mathrm{PCE}(\mathrm{\mathbf{F}_q},\mathrm{\mathbf{F}_S}), 
\end{equation}
and $\mathrm{S}$ is any set of images. 

The Algorithm 1 shows the procedure we used for PRNU subset SCI. More information about the scenarios will be given in Section \ref{sec:results}.

\begin{figure}[!htb]
\centering
\includegraphics[trim={0 1cm 0 0},clip,scale=.6]{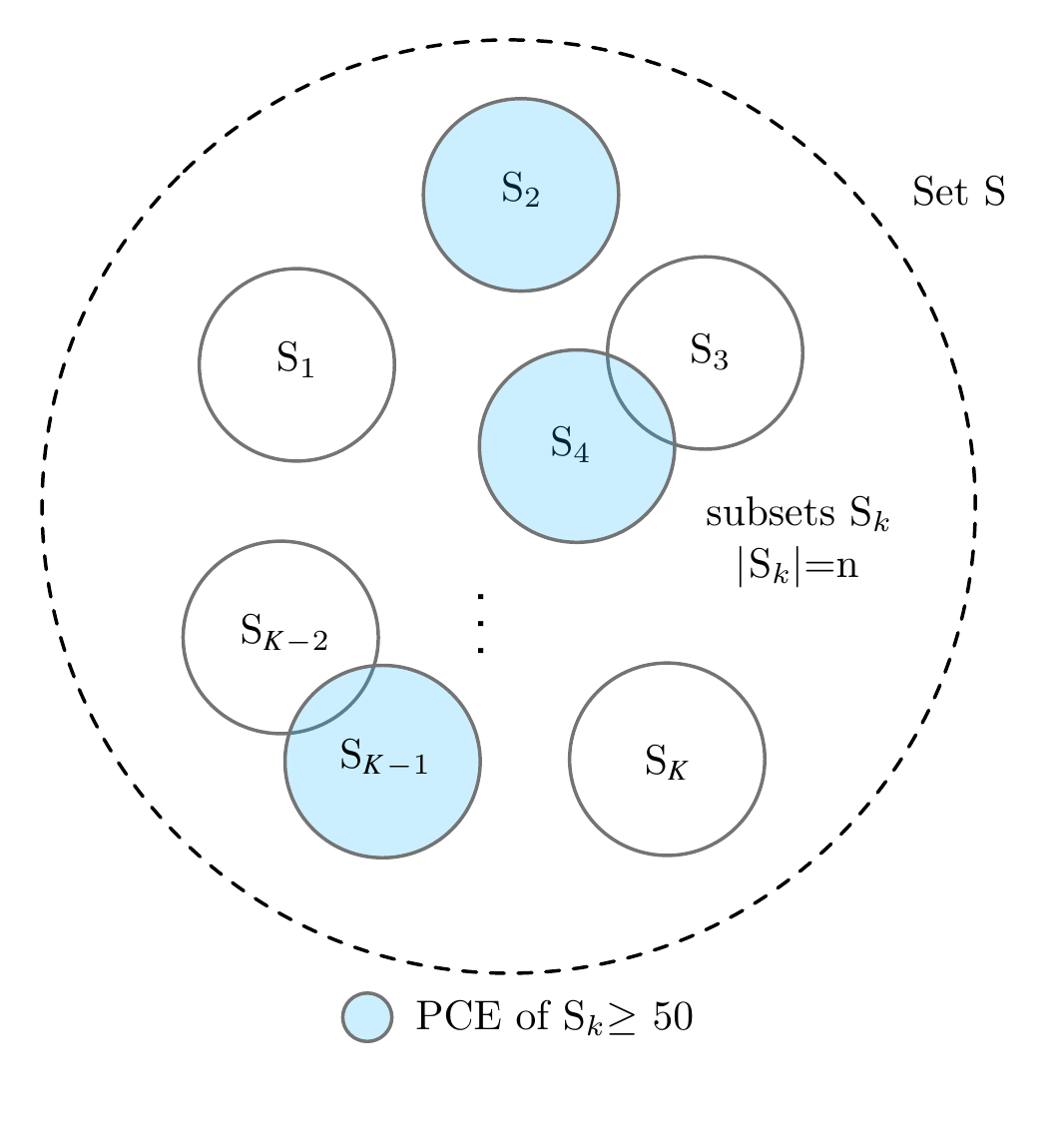}
\caption{Image subsets}
\label{fig:subsets}
\end{figure}

One approach for the attribution subsets of PM-images to their source cameras would be to generate PRNU fingerprints for all possible combinations in a given image dataset ($N$ images), which would consist of $K=2^N-1$ fingerprints. Assuming there were $N=30$ images, a total of $1.07\times 10^9$ fingerprints would be needed. On the other hand, if we were to reduce the number of combinations, $K$ number can be selected sparsely, e.g. from subsets with length of 5 samples to 20 samples with 5 increments, then the  number of fingerprints would reduce to one fifth of the previous amount, which would be still far from practical. 
We therefore limited the number of fingerprints to 100 in each experiment, by limiting the number of subsets to 100 for each 4 subset size in all scenarios we will present. 

In Figure \ref{fig:subsets}, a diagram outlining the usage of subsets is given. As shown in the figure, each subset $\mathrm{S}_k$ is populated by randomly sampling the whole set $\mathrm{S}$ with the subset size, $n$. These subsets are allowed to overlap, however each \textcolor{black}{one} is unique. Such that, there are $K=100$ subsets, each having cardinality $n$, where $n=5,10,15,20$, and any subset is chosen to not be identical with any other subset, which can be formally stated for each $n$ value, as:

\begin{equation}
    \mathrm{S}_i \neq \mathrm{S}_j \Rightarrow \bigm| \mathrm{S}_i \cap \mathrm{S}_j\bigm| < n 
    \label{eq:subsetcondition}
\end{equation}

where $\mathrm{S}_i$ and $\mathrm{S}_j$ are any two subset populated within the loop in the algorithm, \textcolor{black}{with $|\mathrm{S}_i| = |\mathrm{S}_i| = n$}. If the PCE of any subset is found over the threshold $\tau$, for example the PCE of $i$th subset $\mathrm{S}_i > \tau$, the content (list of image paths) of the subset is then added to the $p$th fusion set, $\Phi_p$. When the loop ends, the PCE of $\Phi_p$ is used to calculate the PRNU similarity of the fusion set populated within the loop. \textcolor{black}{$\tau$ is set to $50$ along this paper as the PRNU similarity threshold in terms of PCE as discussed as the lower end in \cite{Goljan2009a}}

\section{Experimental Setup \& Results}
\label{sec:results}

In this section we describe the environment of our experiments for the proposed approach to source camera attribution of PM-image subsets, starting with the creation of the PM database in Section \ref{subsection:thedatabase}, then we explore two main scenarios an analyst may face for SCI on PM-image sets. The first one (in Section \ref{subsection:sc1}) is the homogeneous scenario, where the PM-images are captured only from the known ``query camera'', which can be presumed as the easiest scenario for any SCI task. Followed by the second scenario in Section \ref{subsection:sc2}, the heterogeneous scenario, in which the analyst has to find a link between \textcolor{black}{incriminating} images from a query camera within a set of images including images from an unknown camera, denoted as ``unknown camera''. 

\subsection{Creating a PM-image dataset} 
\label{subsection:thedatabase}

In this study, the dataset we used is based on the Realistic Tampering dataset in \cite{Korus2016TIFS,Korus2016WIFS}, which consists of pristine images along with their manipulated version for four different cameras, each having 55 pristine images with single resolution, which is $1920 \times 1080$ pixels without meta-data fields.

From these pristine images, we randomly selected a list (file names) of 25 images to estimate the camera PRNU fingerprint. The list of the remaining 30 images were reserved for tests. 

The PM implementation we targeted produces only gray-scale images and trims each image by 7 pixels from each side (the trim size is the size of patch window size minus one, where the patch window size is $8\times8$). It is possible to overcome the gray-scale limitation, but we preferred to execute the method as it was in the original paper. As the size and color differences between images might influence our study, we produced two versions for each image: a) PM version, b) non-PM version. Naming convention for these versions are as the follows; ``out-pm-before-file.ext'' for the non-PM version, and ``out-pm-after-file.ext'' for the PM version. The interested readers can find links to the dataset in: \href{https://github.com/akarakucuk/2019_PM_SCI_DATA/}{github.com/akarakucuk/2019\_PM\_SCI\_DATA/}.

The non-PM images were generated by applying the very same crop settings and color conversion. These images were then used to generate the query camera's PRNU fingerprint estimate, $\mathrm{\mathbf{F}_q}$ using the selected list mentioned earlier.

This way, we have a PM version for each non-PM image,
which are both saved once without compression, and allowed us to be able to compare and evaluate the manipulation caused by PM in terms of the manipulated pixel's rate (MPR) and the image quality (PSNR). Some example images are given in Figure \ref{fig:examplePMimages}.

We would like to highlight that, out of the images we mentioned we would use for evaluations of PM-images, there were a few images (1 for A57, 3 for D90), which were still identifiable by the PRNU based SCI. At a first glance, inclusion of these images in our evaluations might be more realistic. \textcolor{black}{However, such images should be simply filtered-out by running individual images through the conventional PRNU based SCI or by running the proposed algorithm by setting $n$ to $1$. Including these images can be problematic in certain scenarios, for example in Scenario \#2 in the Section \ref{subsection:sc2}, the PRNU similarity of subsets dominated by non-matching images could also be lifted over the decision threshold, which could in turn produce a lower performance. By the same token, should we choose to include them in our evaluations in Scenario \#1, Section \ref{subsection:sc1}, it could have served as make-up to our proposed methods' advantage and produce higher performance}. Therefore, we opted not to make use of these PM-images when the performance of the proposed method was evaluated anytime $\mathrm{\mathbf{F}_q}$ corresponds to these cameras. Please note that these 4 individual images are included if the camera PRNU fingerprint estimate $\mathrm{\mathbf{F}_q}$ was not either A57 or D90. 

\begin{figure}[!h]
\centering
    \begin{subfigure}{.45\textwidth}
    \includegraphics[trim = 9px 20px 35px 0px, clip, scale=.44]{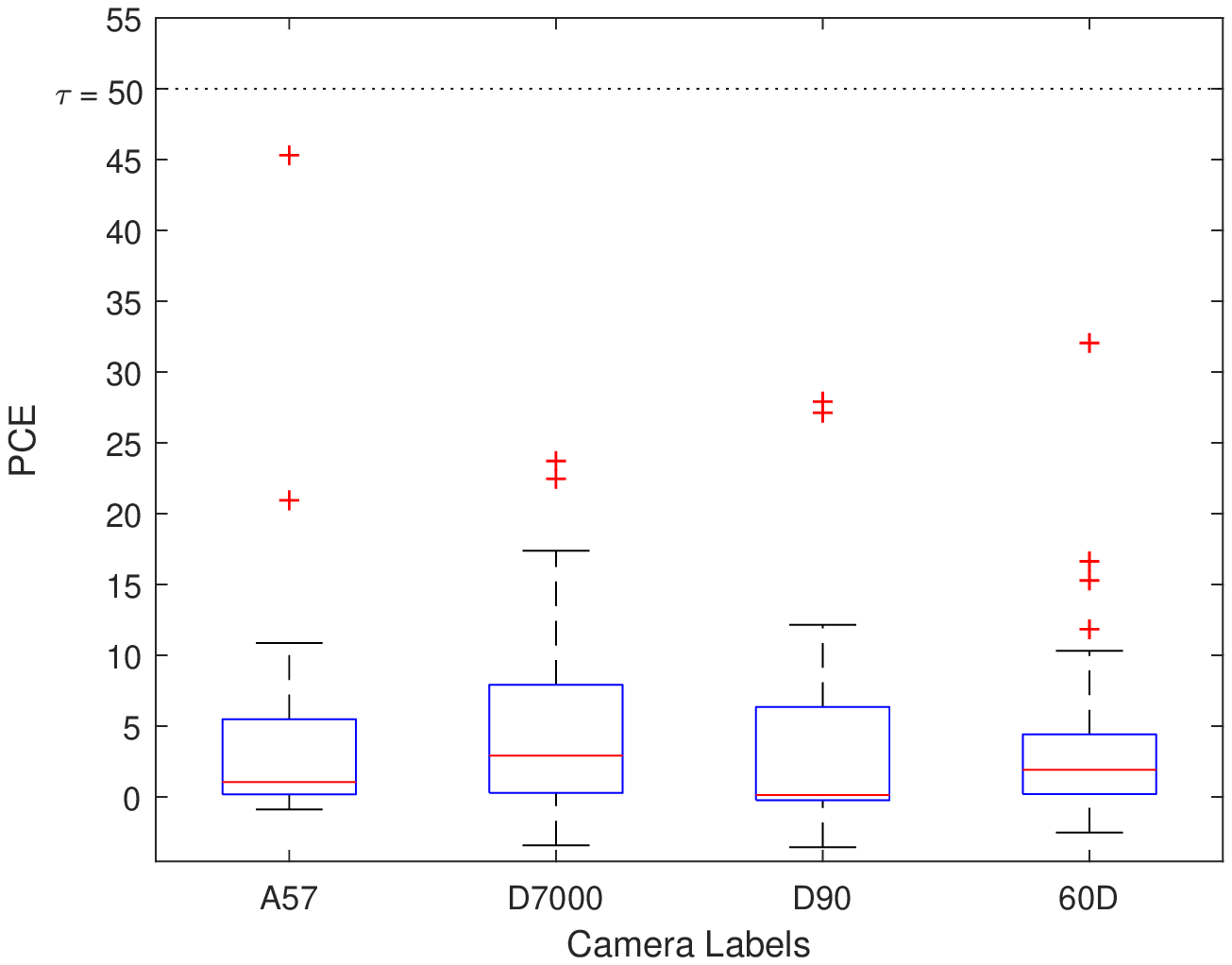} 
    \caption{PCE distribution per camera}
    \label{subfig:initialPMPCEs}
    \end{subfigure}
     \quad
      \begin{subfigure}{.45\textwidth}
    \includegraphics[trim = 9px 20px 35px 0px, clip, scale=.44]{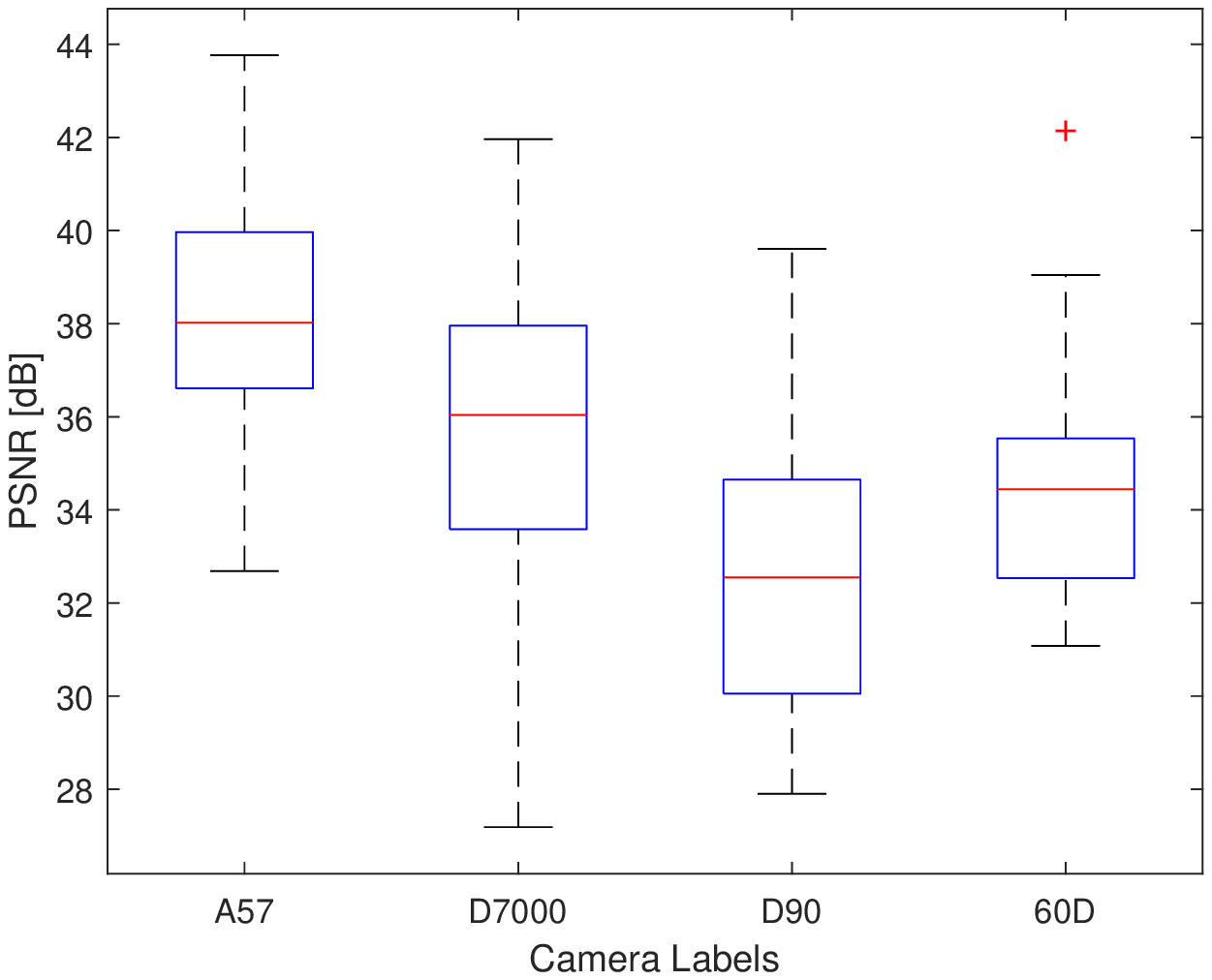}
    \caption{PCE distribution per camera}
    \label{subfig:initialPSNR}
    \end{subfigure}

     \caption{Distributions of PRNU similarity and the image quality of each image in the PM database per camera. In the left the PRNU similarity in terms of PCE and in the right, the image quality metric, PSNR in terms of dB were given. }
     \label{fig:initialPCEandPSNR}
\end{figure}

Readers are referred to Table \ref{tbl:dataset} and its accompanying figure, Fig. \ref{fig:initialPCEandPSNR} to have an understanding regarding the initial state of the dataset. The referred table and the figure show the values of initial PCE, PSNR and Manipulated-Pixel-Rate (MPR) between the PM and the non-PM image versions. The latter shows the percentage of changed pixels of each image pair, and calculated simply by,
\begin{equation}
    \textrm{MPR [\%]}=100 \times \frac{1}{{R} \times {C}} \sum_{r=1}^{{R}}\sum_{c=1}^{{C}}\textrm{sign}({|\mathrm{\mathbf{L}}(r,c)-\mathrm{\mathbf{L}}_{pm}(r,c)|})
    \label{eq:mpr}
\end{equation}

while ignoring signum function $\textrm{sign}()$ at value 0. Here, $\mathrm{\mathbf{L}}$ denotes the non-PM version, $\mathrm{\mathbf{L}}_{pm}$ denotes the PM version of an image. ${R}$ and ${C}$ are the width and height values in pixels. Also in the table are the camera labels that were used to name the cameras. The values given in Table \ref{tbl:dataset} are median values, and they show the manipulation efficiency of the PM, when compared to non-PM values in terms of PCE. For all cameras and images, the observed PCE values were found very low, with high PSNR and MPR rates.

\begin{singlespace}
\tabcolsep=0.08cm
\begin{table}
\centering
\caption{Properties of the Patch-Match Image Data-set. MPR stands for Manipulated Pixel Rate, which is the percentage of pixels that have changed after application of Patch-Match. The metrics shown in the table are median values.}
\begin{tabular}{cc;{1pt/1pt}c;{1pt/1pt}cc;{1pt/1pt}cc} \hline
\multicolumn{2}{c;{1pt/1pt}}{ Camera } & Number of & \multicolumn{2}{c;{1pt/1pt}}{ PCE } & PSNR & MPR  \\
Make  &	Label	   &Images     & non-PM & PM    & [dB]  &  [\%]  \\ 
\hline \hline
Sony    &	A57     & 29	& 1963 & 1.05 &  38   &  85 \\
Nikon   &	D7000   & 30	& 888  & 2.92 &  36   &  78 \\
Nikon	&	D90     & 27 	& 1231 & 0.14 &  33   &  86 \\
Canon	&	60D     & 30	& 1289 & 1.91 &  34   &  86 \\ \hline
\end{tabular}
\label{tbl:dataset}
\end{table}
\end{singlespace}

\subsection{\textbf{Scenario \#1}: Performance on Homogeneous Image Subsets}
\label{subsection:sc1}

In this scenario, we evaluate the performance on a case where the storage media contains PM-images from only the query camera, which represents the most hygienic scenario an analyst can face face in any circumstance. 
Please recall that these 100 image subsets were populated randomly, each including 5 to 20 ($n$) image to reasonably increase the recall rate of the images as mentioned in Section \ref{section:theproposedmethod}.
\begin{singlespace}
\tabcolsep=0.08cm
\begin{table}\centering
\caption{Scenario \#1, Median and Average PCE values of PM-image subsets, with the exception of n=1$^*$ which does not from a subset and provided only as a reference. Values above the detection threshold are emphasized in bold characters.}
\begin{tabular}{c;{1pt/1pt}ccccc;{1pt/1pt}ccccc}\hline
Camera & \multicolumn{5}{c;{1pt/1pt}}{ Median PCE } &\multicolumn{5}{c}{ Maximum PCE } \\
Label & $n$=$1^*$& $n$=5 & $n$=10 & $n$=15 & $n$=20 & $n$=$1^*$& $n$=5 & $n$=10 & $n$=15 & $n$=20\\ \hline \hline
A57 & 1.1 & 14.3     & 36.0      & 51.6  & 77.0  & 45.3 & \textbf{82.7} & \textbf{80.6} & \textbf{94.1} & \textbf{109.3}  \\
D7000 & 2.9 & 11.5   & 18.2      & 36.0  & 45.9 & 23.7 & {48.1} & \textbf{63.2} & \textbf{71.9} & \textbf{83.9} \\
D90 & 0.1 & 3.7      & 7.4       & 9.7   & 12.6 & 27.9 & \textbf{55.7} & {43.8} & {35.9} & {27.9}  \\
60D & 1.9 & 13.1     & 22.3      & 33.3  & 47.2  & 32.1 & \textbf{54.2} & \textbf{52.5} & \textbf{80.6} & \textbf{79.9}  \\
\hline
\end{tabular}
\label{tbl:only-query}
\end{table}
\end{singlespace}

The PCE values are given in Table \ref{tbl:only-query} and indicate a few cases where the set attribution approach may fail. Specifically, the D90 has an unexpected result, where the performance gets worse with elevated $n$ values. One possible explanation could be based on the initial median values of PM-images for this camera (Table \ref{tbl:dataset}) which is, $0.14$ for PCE. This is an eighth of its closest performer, A57 camera in the same test, which was $1.05$. The distribution of the values is also similar in terms of both PCE and PSNR distribution for this particular camera, as it can be seen in Fig. \ref{fig:initialPCEandPSNR}. This indicates that the majority of images from D90 are more heavily affected by PM, the performance of D90, with only one subset with $n$=5 was above the PCE threshold. We'll compare this and other results more closely through the coming table, Table \ref{tbl:fusionSC1}.

Table \ref{tbl:fusionSC1} shows the results from the fusion set, broken down by each camera label and subset size ($n$) the test was conducted on. 
The fusion sets were populated by setting $\mathrm{S_\beta}$ constantly to $\varnothing$ in Algorithm 1, in order to be in line with the Scenario \#1. In Table \ref{tbl:fusionSC1}, $|\mathrm{\Phi}_p|$ denotes the total number of images i.e. cardinality, of each fusion set and their PCE values, s.t. PCE of $\mathrm{\Phi}_p$. PCE values here represent the PRNU similarity when all images in each fusion set were incorporated. The ``Recall'' value were also given in this table, which shows the \textcolor{black}{recall} rate of PM-images that belong to the labeled device populated into the fusion set, formally given by:
\begin{equation}
\textrm{Recall [\%]} = 100 \times \frac{|\Phi_p|}{|\mathrm{S_\alpha}|}.
\label{eq:recall}
\end{equation}

There are missing fusion sets, denoted with ``--'' in the table. For example on D7000, there was no fusion set for $n$=5, meaning none of the PM-image subsets for this camera with this subset size were reached over PCE value threshold $\tau$. In addition, D90 also indicates that there is only one subset of PM-images brought together a combined PRNU noise pattern that matched with the camera's PRNU fingerprint estimate.
The results from the remaining sets ($n$=10 to $n$=20) through this scenario confirm this one-off situation as none of them provided similar rates.

\begin{singlespace}
\tabcolsep=0.08cm
\begin{table}
\centering
\caption{Scenario \#1: Details of fusion sets. ``--'' represents cases without an outcome. PCE column represents PCE of $\mathrm{\Phi}_p$.}
\begin{tabular}{c;{1pt/1pt}c;{1pt/1pt}cc;{1pt/1pt}cc} \hline
Camera &    & \multicolumn{2}{c;{1pt/1pt}}{Fusion Set} & \multicolumn{2}{c}{Recall} \\
Label	                & $n$         & $|\Phi_p|$    &   PCE      & Value & [\%] \\ \hline
\multirow{4}{*}{A57}  	& 5      	&   22                 &   100.2    & 22/29 & 76 \\
						& 10        &   29                 &   100.0    & 29/29 & 100\\
						& 15    	&   29                 &   100.0    & 29/29 & 100 \\
						& 20     	&   29                 &   100.0    & 29/29 & 100 \\ \hdashline[1pt/1pt]
    				            
\multirow{3}{*}{D7000}  & 5        &   --	               &   --     &  -- & -- \\
                        & 10        &   16	               &   83.2     & 16/30 & 53\\
						& 15        &   30	               &   67.3     & 30/30 & 100 \\
						& 20        &   30	               &   67.3     & 30/30 & 100 \\ \hdashline[1pt/1pt]
\multirow{3}{*}{D90}     & 5         &   5                  &   55.7     & 5/27 & 18\\ 
				        & 10        &    --	               &    --     & --&--\\
				        & 15        &    --	               &    --     & --&--\\
				        & 20        &    --	               &    --     & --&--\\ \hdashline[1pt/1pt]
\multirow{4}{*}{60D} 	& 5	    	&   9	                &   82.8    & 9/30 & 30 \\ 
						& 10     	&   24	                &   81.7    & 24/30 & 80\\
						& 15    	&   30	                &   68.7    & 30/30 & 100\\
						& 20    	&   30	                &   68.7    & 30/30 & 100\\ \hline
\end{tabular}
\label{tbl:fusionSC1}
\end{table}
\end{singlespace}

It is evident in this table that the proposed method has \textcolor{black}{reached complete recall} with $n$=15 on all the three cameras, however, in D90, only 18\% of the PM-images were covered with a PCE value of 55.7 on $n$=5. In this table, it is also evident that the increased cardinality values meet with decrease in PRNU similarity in terms of PCE. The most notable is D7000, where the PRNU similarity drops from 83.2 to 67.3 when the fusion set reached from about half of the PM-images to all of such images. Similar tendencies are evident for different cameras as well. The exception is also the camera D7000, where none of the 100 subsets with cardinality 5 ($n$=5) did not reach or exceed the PCE threshold. This may indicate a need to increase the  number of randomly selected subsets.

\begin{figure}[!h]
\centering
    \begin{subfigure}{0.45\textwidth}
    \includegraphics[trim = 9px 20px 35px 0px, clip,     scale=.44]{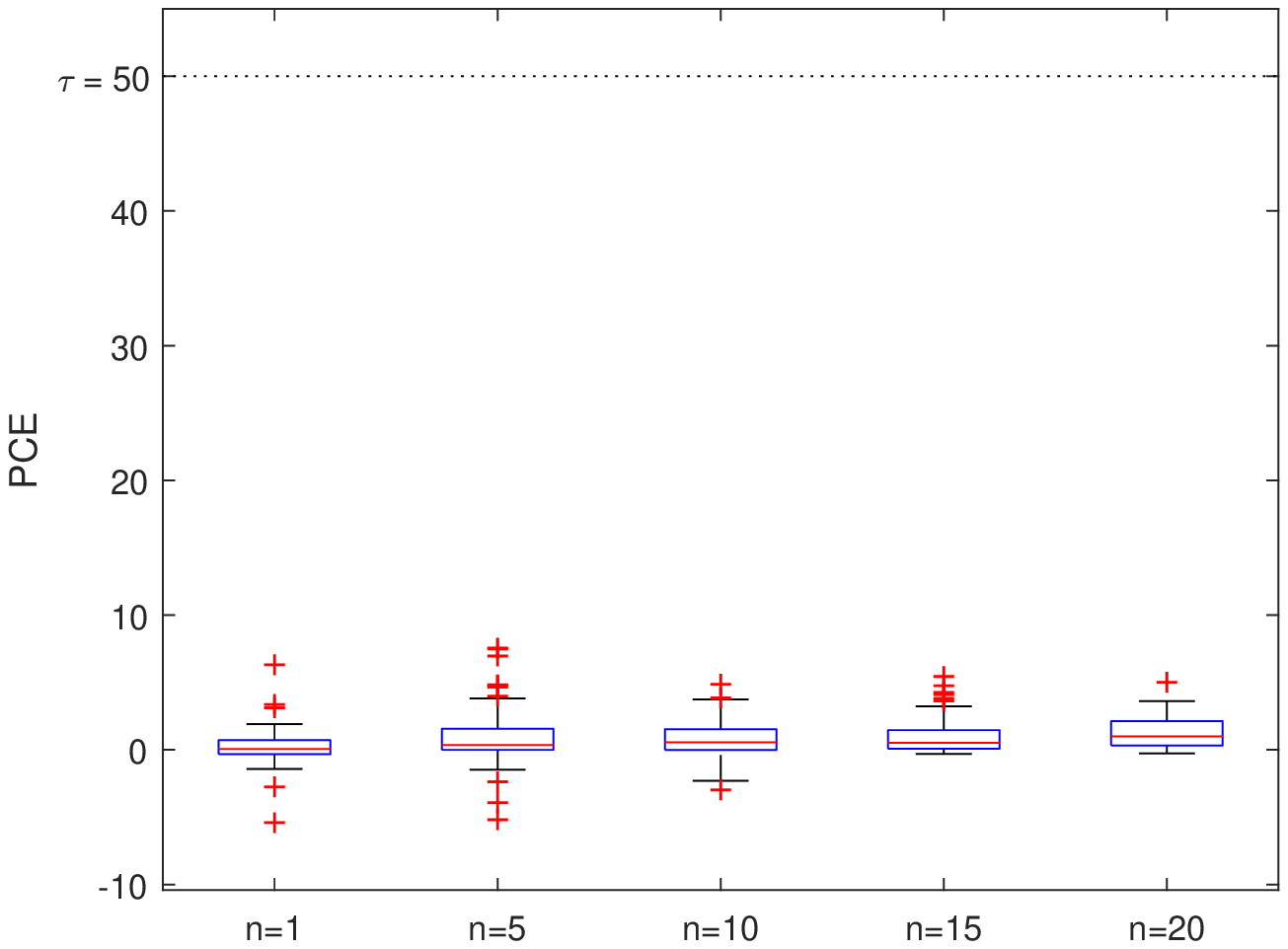} 
    \caption{Non-matching PM-images.}
    \label{subfig:NonMatchingGroups}
    \end{subfigure}
     \quad
     \begin{subfigure}{0.45\textwidth}
   \includegraphics[trim = 9px 20px 35px 0px, clip, scale=.44]{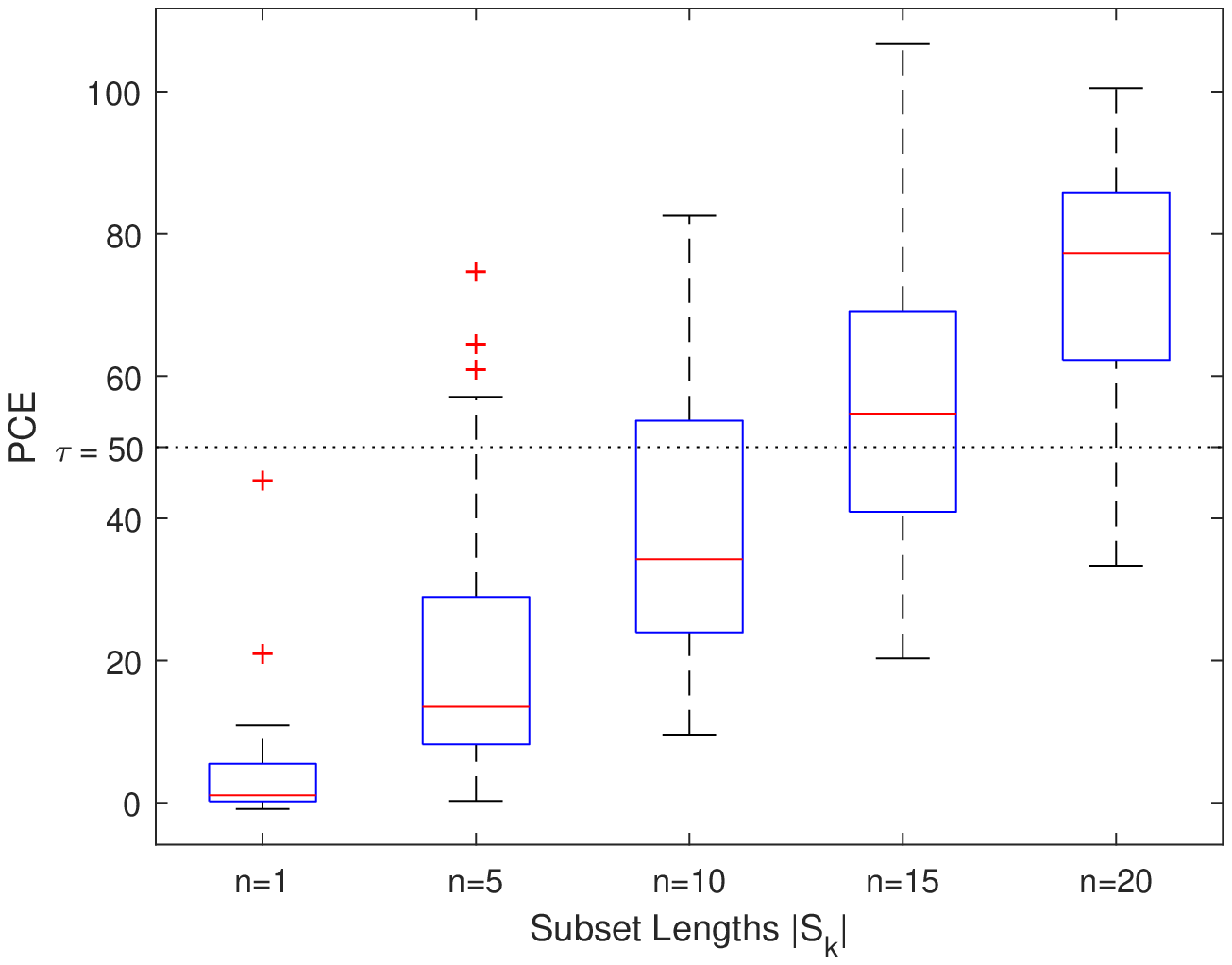}
    \caption{Matching PM-images.}
    \label{subfig:MatchingGroups}
    \end{subfigure}
\caption{In (a) non-matching case (PM-image subsets from Canon 60D and PRNU fingerprint from Sony A57) and in (b), matching case (PM-image subsets from Sony A57 and PRNU fingerprint from Sony A57). The PRNU similarity values of these subsets were obtained against $\mathrm{\mathrm{\mathbf{F}_q}}$ of Sony A57. In both figures, boxes on the far left are for the subset size of $n$=1$^*$ and only given to serve as a reference PCE distribution for per-image statistics.}
     \label{fig:groupsforallN}
   \end{figure}

On a different note, the PRNU similarity of the PM-images from non-matching devices were also observed to see if such subsets can reach or exceed the threshold. As expected, there was none. The results regarding the non-matching and matching camera PM-images for a few example cameras were plotted in Fig. \ref{fig:groupsforallN}. More details regarding the results from the non-matching cases were omitted as they all give the same outcome.

\subsection{\textbf{Scenario \#2}: Performance on Heterogeneous Image Sets from Camera Pairs}  \label{subsection:sc2}

In this scenario, to evaluate the performance of the proposed method when images from an unknown origin are also present along with PM-images of the query camera in the storage media.

To do so, we'd like to start by elaborating the concept of Case IDs. A Case ID simply refers to a specific pair of source cameras as shown in Table \ref{tbl:caseids}. For example, the Case ID 1 represents a pair when the storage media has images from both camera A57 and D7000. When all images (59) from these two cameras, which we denote by $\mathrm{S_\Sigma}$, were merged, the PRNU similarity is found as 32.5 as listed on the first row under the PCE column, s.t. PCE of $S_\Sigma$ for Case 1, in the same table. Please recall, $\mathrm{S_\Sigma}$ term denotes a collection of images from two different cameras. The analyst is assumed to have knowledge of the PRNU fingerprint he gathered from a separate set of 25 pristine images only from the query camera. 
Also recall that the storage media the analyst received only has PM-images from the same query camera, 
whereas the images from the unknown camera has 50\% chances of being a PM-image, \textcolor{black}{to be avoid of any form of bias}. In Table \ref{tbl:caseids}, initial PRNU similarities observed in terms of PCE of the listed cameras were also given.

To remind the readers about the PRNU Subset and Fusion SCI algorithm, we would like to refer once again to Algorithm 1. In the algorithm, $p$ represents a descriptor for camera pairs, which is an integer ranging from 1 to 12 with subset sizes $n$ starting from 5 to 20, with 5 image increments. Same number of random subsets ($K$=100) for each length per each pair were selected, as indicated in the algorithm with subscript $k$. Thus, the range of these values are $p$=$1,2,...,12$, $n$=$5,10,15,20$ and $k$=$1,2,...,100$, 
as indicated in Table \ref{tbl:caseids}. Please also note that the total number of images ($|\mathrm{S_\Sigma} |=|\mathrm{S_\alpha} \cup \mathrm{S_\beta}|$) vary slightly on different pairs. 

The initial state of the PRNU similarities observed from the table indicates that in two cases (Case 3 and 5), the analyst might find it especially hard to distinguish images from both cameras if they were naively combined and could end up with 50\% probability of mislabeling images from the unknown camera. The proposed method, on the hand, can achieve up to 100\% precision on 6 different cases as shown in Table \ref{tbl:fusionscenariomix}.

\begin{singlespace}
\tabcolsep=0.08cm
\begin{table}\centering
\caption{Scenario \#2: Properties of Cases. The right-most column represents PCE value when all images from both cameras are combined. $|\mathrm{S_\alpha}|$ and $|\mathrm{S_\beta}|$ represents number of images in the PM-image set from the query camera and mixed type of images (PM-image + original) from the unknown camera listed in each row.}
\begin{tabular}{c;{1pt/1pt}cc;{1pt/1pt}cc;{1pt/1pt}c} \hline 
Case & \multicolumn{2}{c;{1pt/1pt}}{ Camera Labels } & \multicolumn{2}{c;{1pt/1pt}}{} &  PCE\\
ID& {Query} & Unknown &   $|\mathrm{S_\alpha}|$ &  $|\mathrm{S_\mathrm{\beta}}|$   &  of $\mathrm{S_\Sigma}$\\ 
\hline \hline
1 & A57 & D7000 & 29 & 30   & 32.5 \\ 
2 & A57 & D90 & 29 & 30     & 30.7 \\ 
3 & A57 & 60D & 29 & 30     & \textbf{78.3} \\ 
4 & D7000 & A57 & 30 & 30   & 25.6 \\ 
5 & D7000 & D90 & 30 & 30   & \textbf{57.0} \\ 
6 & D7000 & 60D & 30 & 30   & 23.2 \\ 
7 & D90 & A57 & 27 & 30     & 12.1 \\ 
8 & D90 & D7000 & 27 & 30   & 6.2 \\ 
9 & D90 & 60D & 27 & 30     & 4.3 \\ 
10 & 60D & A57 & 30 & 30    & 45.9 \\ 
11 & 60D & D7000 & 30 & 30  & 26.0 \\ 
12 & 60D & D90 & 30 & 30    & 40.1 \\ \hline 
\end{tabular}
\\
\vspace{2pt}
\label{tbl:caseids}
\end{table}
\end{singlespace}

For each case listed in the Table \ref{tbl:caseids}, PRNU noise from 100 randomly populated subsets, each having the subset sizes of $n$ = $5,10,15,20$ number of images were used to estimate a PRNU fingerprint and correlated with the query PRNU fingerprint $\mathrm{\mathrm{\mathbf{F}_q}}$. The median and maximum PCE values from these subsets were shown in Table \ref{tbl:full-mixedMEDIAN}.
\begin{singlespace}
\tabcolsep=0.08cm
\begin{table}\centering
\caption{Scenario \#2, PRNU values above the detection threshold are emphasized in bold characters, with the exception of $n$=1$^*$ which does not constitute a subset and provided only as reference. Please note that the maximum values in the n=1$^*$ column are the same as Scenario \#1. Values above the detection threshold are emphasized in bold characters.}
\begin{tabular}{c;{1pt/1pt}ccccc;{1pt/1pt}ccccc} \hline
Case &	\multicolumn{5}{c;{1pt/1pt}}{Median PCE} &	\multicolumn{5}{c}{Maximum PCE}	\\ 
ID  & $n$=$1^*$   &$n$=$5$&$n$=$10$&$n$=$15$&$n$=$20$  & $n$=$1^*$   &$n$=$5$&$n$=$10$&$n$=$15$&$n$=$20$ \\ \hline \hline
{1} & 0.2       &2.1&5.1&5.6&11.2&   45.3 &41.8&\textbf{52.7}&33.9&39.7 \\
{2} & 0.1       &1.5&6.0&7.8&11.0&   45.3 &\textbf{65.1}&37&41.3&44.6  \\
{3} & 0.5       &6.0&14.8&19.8&23.8& 45.3 &47.7&\textbf{59.3}&\textbf{66.5}&\textbf{74.6} \\
4   & 0.1       &3.2&5.5&6.6&10.5&   23.7 &36.6&32.7&49.5&39.2  \\
{5} & 0.6       &5.3&11.9&14.1&19.5& 23.7 &35.5&\textbf{58.2}&\textbf{63.4}&\textbf{54.3}  \\
{6} & 0.3       &2.7&3.8&6.8&6.2&    23.7 &24.9&33.7&37.6&\textbf{50.2}  \\
7   & 0.2       &0.9&2.6&4.4&4.4&    27.9 &19.5&27.3&21.1&24.2  \\
8   & 0.0       &0.8&1.5&2.1&3.5&    27.9 &34.4&19.4&19.7&33.3 \\
9   & 0.0       &0.3&0.7&2.0&2.4&    27.9 &34.7&30.2&31&21.6 \\
{10}& 1.0       &4.2&7.3&11.8&15.3&  32.1 &39.3&45.3&43.6&\textbf{57.2}  \\
11  & 0.7       &2.4&3.9&7.0&9.1&    32.1 &25.4&32.6&27.1&35.6 \\
12  & 0.2       &2.6&6.8&10.0&13.6&  32.1 &31.2&44&39.1&45.6 \\ \hline
\end{tabular}
\label{tbl:full-mixedMEDIAN}
\end{table}
\end{singlespace}

In Table \ref{tbl:full-mixedMEDIAN}, values from all subsets are given and values over the threshold are emphasized in bold characters. This table shows the PRNU similarity of the incorporated subsets in terms of PCE values. In many cases, the subsets did not reach or exceed the threshold, but any subset having a PCE value over the threshold were incorporated for each $n$ and $p$ which forms the fusion set $\mathrm{\Phi}_p$, and all images in this set were then used to re-calculate the final PCE of $\mathrm{\Phi}_p$. The results and the number of images in the fusion set along with the Precision rates for each case are shown in Table \ref{tbl:fusionscenariomix}. Precision value shows the ratio of PM-images from the query camera in the fusion set, and in terms of percentage, can be formally given as:

\begin{equation}
\mathrm{Precision} [\%] = 100 \times \frac{|\Phi_p \cap \mathrm{S_\alpha}|}{|\mathrm{\Phi}_p|}
\label{eq:precision}
\end{equation}

\begin{singlespace}
\tabcolsep=0.08cm
\begin{table}
\centering
\caption{Scenario \#2: Details of the fusion sets.  Cases without an outcome are represented with ``-–'' mark.}
\begin{tabular}{c;{1pt/1pt}c;{1pt/1pt}cc;{1pt/1pt}cc} \hline
Case &  & \multicolumn{2}{c;{1pt/1pt}}{Fusion Set} & \multicolumn{2}{c}{Precision} \\
ID	                & $n$ & $|\Phi_p|$ & PCE 	& Value & [\%] \\ \hline
	
1 					& 10    &10     & 52.7		&9/10 & 90\\ \hdashline[1pt/1pt]
2 					& 5 	&5      & 65.1 		& 5/5 & 100\\ \hdashline[1pt/1pt]
\multirow{3}{*}{3} 	& 10    &26     & 92.1 		& 18/26& 69\\
 					& 15 	&46		& 92.4      & 26/46& 57\\
 					& 20 	&47		& 90.2      &  26/47&55\\ \hdashline[1pt/1pt]
 				4	& -- 	&--		& --      &  --&--\\ \hdashline[1pt/1pt]
\multirow{3}{*}{5} 	& 10    &10		& 58.2 		& 4/10&40\\
 					& 15    &24		& 77.9 		& 16/24& 67\\
 					& 20    &33		& 64.5 		& 17/33& 52\\ \hdashline[1pt/1pt]
6 					& 20	&20		& 50.2 	    & 14/20& 70\\ \hdashline[1pt/1pt]
 				7	& -- 	&--		& --      &  --&--\\ \hdashline[1pt/1pt]
 				8	& -- 	&--		& --      &  --&--\\ \hdashline[1pt/1pt]
 				9	& -- 	&--		& --      &  --&--\\ \hdashline[1pt/1pt]
10 					& 20	&32		& 63.3 	    & 18/33& 59\\ \hline
11	& -- 	&--		& --      &  --&--\\ \hdashline[1pt/1pt]
12	& -- 	&--		& --      &  --&--\\  \hline
\end{tabular}
\label{tbl:fusionscenariomix}
\end{table}
\end{singlespace}

In Table \ref{tbl:fusionscenariomix}, there are only two cases, Case 3 and 5 that produced a fusion set for plural number of subset sizes, namely, for $n$=10,15 and 20. This calls for a further elimination for these fusion sets, which can be done by finding the intersection of fusion sets for all available $n$ values. This produces $|\mathrm{\Phi}_{p=3}|= 18$ which has 14 PM-image from the query camera, and improves the precision to 78\%, which is higher than all individual fusion sets, which have a maximum precision of 69\%. Another such case is Case 5, where the precision, with this type of elimination yields with 4/6 and 67\%, which was the maximum value in this case when $n$=15 in the same table. This type of elimination can find its use for highly critical tasks.

In Table \ref{tbl:fusionNoverview}, a summary showing the overall performance of the algorithm for this scenario is given, broken down for each $n$ values. Here, the column labeled as ``C'' indicates the cases where the algorithm produced any subset having a PRNU similarity over the threshold in terms of PCE value. Values indicated under column ``Average $\Phi$'' shows the average length of fusion sets for available cases. 

There are three metrics, namely, ``T.Precision'',  ``T. Recall'' and ``Selection''. \textcolor{black}{The last metric shows the representation rate of images in fusion sets, which is calculated by dividing the total number of images in the fusion sets to the total number of images in the represented cases}, which can be written by using $\sum_{p \in C}{}$ operator which is used to restrict the results to cases having a fusion set, as:

\begin{equation}
    \label{eq:selection}
\textrm{Selection [\%]} = 100 \times \frac{\sum_{p \in \mathrm{C}}{} |\mathrm{\Phi}_p|}{\sum_{p \in \mathrm{C}} |\mathrm{S_\Sigma}|}
\end{equation}

\textcolor{black}{The other values in the Table \ref{tbl:fusionNoverview}, ``T.Precision'' and ``T.Recall'' denotes  for Total Precision and Total Recall and indicates how well the proposed method worked in terms of accuracy and sensitivity}. These were also calculated by summing over $p\in \mathrm{C}$, given by:

\begin{equation}
    \label{eq:trecall}
\textrm{Total Recall} [\%] = 100 \times \frac{\sum_{p \in \mathrm{C}}{} |\mathrm{\Phi}_p \cap \mathrm{S_\alpha}|}{\sum_{p \in \mathrm{C}} |\mathrm{S_\alpha}|}
\end{equation}

\begin{equation}
    \label{eq:tprecisionl}
    \textrm{Total Precision [\%]} = 100 \times \frac{\sum_{p \in \mathrm{C}}{} |\mathrm{\Phi}_p \cap \mathrm{S_\alpha}|}{\sum_{p \in \mathrm{C}} |\mathrm{\Phi}_p|}.
\end{equation}
\begin{singlespace}
\tabcolsep=0.08cm
\begin{table}
\centering
\caption{Scenario \#2: Performance of PRNU based SCI w.r.t. each subset length $n$.}
\begin{tabular}{c;{1pt/1pt}c;{1pt/1pt}cr;{1pt/1pt}cr;{1pt/1pt}cr;{1pt/1pt}c} \hline
 &  Cases & \multicolumn{2}{c;{1pt/1pt}}{T.Precision}  & \multicolumn{2}{c;{1pt/1pt}}{T.Recall} & \multicolumn{2}{c;{1pt/1pt}}{Selection}          & Average\\ 
$n$ & C  & Value & [\%]  & Value & [\%] & Value & [\%]          &of $|\mathrm{\Phi}|$  \\ \hline \hline
5	& 2 	& 5/5       &   100 &     5/29  &   17	&   5/59        &    8		& 5 \\
10	& 1,3,5 	& 31/46     &   67  &    31/88  &   35	&   46/178      &   26		& 15 \\
15	& 3,5 	& 42/70     &   60  &    42/59  &   71	&   70/119      &   59		& 35 \\
20	& 3,5,6,10 	& 76/132    &   58  &   76/119  &   64	&   132/239     &   55		& 33 \\ \hline
\end{tabular}
\\
\vspace{2pt}
\label{tbl:fusionNoverview}
\end{table}
\end{singlespace}

The results in Table \ref{tbl:fusionNoverview} shows that the most cases were reported with the subset size $n$=20, however it also had the lowest Total Precision value. The best Total Precision was reported when the subset size was the smallest, $n$=5, which was expected. Because, by allowing more images in the subsets (with PCE value over the $\tau$), the chances of including images of unknown origin also increase. This may indicate that increasing the number of subsets $K$ for smaller subsets would be a beneficial trade-off for the analyst. \textcolor{black}{The influence of the number of subsets ($K$ parameter in the Algorithm 1) will be discussed in the next section.}

\begin{singlespace}
\tabcolsep=0.08cm
\begin{table}
\centering
\caption{Scenario \#1: The T. Recall rates of fusion sets w.r.t. the subset length ($n$) and the number of subsets ($K$). The columns under percentage symbol are the average recall rates  Eq. \ref{eq:trecall}, and the columns under C denote the number of query cameras producing a fusion set.}
\begin{tabular}{c;{1pt/1pt}ccc;{1pt/1pt}ccc;{1pt/1pt}ccc;{1pt/1pt}ccc} \hline
  & \multicolumn{3}{c;{1pt/1pt}}{$n=5$} & \multicolumn{3}{c;{1pt/1pt}}{$n=10$} & \multicolumn{3}{c;{1pt/1pt}}{$n=15$} & \multicolumn{3}{c}{$n=20$} \\
	              $K$ & $|\Phi_p|$ & [\%]	& C & $|\Phi_p|$ & [\%]& C& $|\Phi_p|$ & [\%]	& C& $|\Phi_p|$ & [\%]& C	 \\ \hline \hline
10	&	5	&	17	&	1	&	15	&	52	&	1	&	20	&	66	&	3	&	29	&	99	&	3\\ 
20	&	7	&	22	&	2	&	25	&	86	&	1	&	26	&	87	&	3	&	30	&	100	&	3\\ 
30	&	9	&	29	&	2	&	29	&	100	&	1	&	26	&	88	&	3	&	30	&	100	&	3\\
40	&	9	&	29	&	2	&	20	&	66	&	2	&	28	&	96	&	3	&	30	&	100	&	3\\
50	&	9	&	29	&	2	&	20	&	66	&	2	&	29	&	99	&	3	&	30	&	100	&	3\\ 
60	&	12	&	41	&	2	&	19	&	63	&	3	&	29	&	99	&	3	&	30	&	100	&	3\\ 
70	&	14	&	47	&	2	&	20	&	67	&	3	&	29	&	99	&	3	&	30	&	100	&	3\\
80	&	14	&	47	&	2	&	22	&	74	&	3	&	29	&	99	&	3	&	30	&	100	&	3\\
90	&	14	&	47	&	2	&	22	&	74	&	3	&	30	&	100	&	3	&	30	&	100	&	3\\ 
100	&	12	&	42	&	3	&	23	&	78	&	3	&	30	&	100	&	3	&	30	&	100	&	3\\ \hline
\end{tabular}
\label{tbl:fusionKbreakdownSC1}
\end{table}
\end{singlespace}

\begin{singlespace}
\tabcolsep=0.08cm
\begin{table}
\centering
\caption{Scenario \#2: The T. Precision rates of fusion sets w.r.t. the subset length ($n$) and the number of subsets ($K$). $n$ and $K$ values without an outcome are represented with ``-–'' mark. The columns under percentage symbol are the average precision rates in Eq. \ref{eq:tprecisionl}, and the columns under C denote the number of cases producing a fusion set.}
\begin{tabular}{c;{1pt/1pt}ccc;{1pt/1pt}ccc;{1pt/1pt}ccc;{1pt/1pt}ccc} \hline
  & \multicolumn{3}{c;{1pt/1pt}}{$n=5$} & \multicolumn{3}{c;{1pt/1pt}}{$n=10$} & \multicolumn{3}{c;{1pt/1pt}}{$n=15$} & \multicolumn{3}{c}{$n=20$} \\
	               $K$ & $|\Phi_p|$ & [\%]	& C & $|\Phi_p|$ & [\%]& C& $|\Phi_p|$ & [\%]	& C& $|\Phi_p|$ & [\%]& C	 \\ \hline \hline
10	&	--	&	--	&	--	&	10	&	40	&	1	&	15	&	73	&	1	&	34	&	59	&	1 \\ 
20	&	--	&	--	&	--	&	10	&	63	&	3	&	15	&	70	&	2	&	41	&	59	&	1 \\ 
30	&	--	&	--	&	--	&	13	&	66	&	3	&	15	&	70	&	2	&	37	&	55	&	2 \\
40	&	--	&	--	&	--	&	13	&	66	&	3	&	15	&	70	&	2	&	37	&	55	&	2 \\
50	&	--	&	--	&	--	&	13	&	66	&	3	&	26	&	61	&	2	&	30	&	58	&	4 \\ 
60	&	--	&	--	&	--	&	13	&	66	&	3	&	26	&	61	&	2	&	30	&	58	&	4 \\ 
70	&	--	&	--	&	--	&	13	&	66	&	3	&	30	&	60	&	2	&	33	&	58	&	4 \\
80	&	--	&	--	&	--	&	13	&	66	&	3	&	33	&	61	&	2	&	33	&	58	&	4 \\
90	&	5	&	100	&	1	&	15	&	67	&	3	&	35	&	60	&	2	&	33	&	58	&	4 \\ 
100	&	5	&	100	&	1	&	15	&	67	&	3	&	35	&	60	&	2	&	33	&	58	&	4 \\ \hline
\end{tabular}
\label{tbl:fusionKbreakdownSC2}
\end{table}
\end{singlespace}

\section{Conclusion}

In this paper, we proposed the first SCI method for images processed with Patch-Match (PM) algorithm. The PM algorithm was originally developed as an image in-painting algorithm, but recently, its use as an attack method against PRNU based SCI was successfully demonstrated. Due to its nature, the algorithm changes many pixels (up to 86\%), which breaks the synchronization of the PRNU noise pattern in images, thus making PRNU based SCI almost obsolete (on 97\% of images we have tested) in identifying individual images through the use of PRNU based SCI method. 

We propose to identify such images \textcolor{black}{using a fixed number of small subsets}, and by using the conventional PRNU similarity metric as a guide to reach to a bigger set, which we call a ``fusion set''. 
The proposed method is evaluated briefly in two scenarios. In the first scenario, the proposed method was tested in a homogeneous setting, where the analyst worked to give an answer if all the incriminating images he received were coming from a suspected, query camera. In this setting, the proposed method is shown to have up to 100\% chance of finding the images on 3 out of 4 source cameras tested. 

In second scenario, the analyst was given a more difficult task, because half of the images the analyst received are coming from an unknown source. In this scenario the proposed method increased the likelihood of correct source identification as well, however there exists quite a few cases (6 out of 12) where there were no conclusion.

\textcolor{black}{As mentioned earlier, to specify an upper bound in computing time we had limited the number of small subsets ($K$) to 100 for all the analysis in this study. Nevertheless, we would like to discuss the influence of $K$ on the analysis in both scenarios using the results we have, in Tables \ref{tbl:fusionKbreakdownSC1} and \ref{tbl:fusionKbreakdownSC2}. In both tables, the results were summarized in terms of $K$, the number of small subsets and $n$, the number of images in the subsets. Regardless of the scenario, increasing the number of $K$ improves the SCI of more cases and query cameras, however the performance rates remain stable.}

We believe the experiments and dataset released along with this study will help to advance the digital forensics related SCI research in the era of software-enriched images. 

In our future studies, we plan to work on identifying traces of Patch-Match on PM-images and evaluate the proposed method outlined in this paper for manipulations with similar nature, such as image in-painting.

\section{Acknowledgements}
The authors want to thank Matthias Kirchner for allowing us to use the Patch-Match code for PRNU de-synchronization. We would also like to thank Pawel Korus for allowing us to re-produce a version Realistic Tampering Database for the purposes of Patch-Match related SCI research.

\bibliography{main}

\end{document}